\documentclass[conference]{IEEEtran}
\usepackage[boxruled]{algorithm2e}
\usepackage{cite}
\usepackage{graphicx}
\usepackage{psfrag}
\usepackage{subfigure}
\usepackage{url}
\usepackage{amsmath}
\usepackage{array}
\usepackage{amssymb}
\usepackage{amsfonts}
\usepackage{graphicx}
\usepackage{epstopdf}
\usepackage{algorithmic}

\newtheorem{lemma}{Lemma}

\newtheorem{definition}{Definition}

\newtheorem{example}{Example}
\newtheorem{note}{Note}
\usepackage{setspace}
\usepackage{amscd}
\usepackage{mathrsfs}
\usepackage{epsfig}
\usepackage{color}
\usepackage{textcomp}
\usepackage{multirow}
\usepackage{threeparttable}
\input{epsf.sty}
\title{Wireless Network Coding for MIMO Two-way Relaying using Latin Rectangles}
\begin{document}

\author{
\authorblockN{Vijayvaradharaj T. Muralidharan and B. Sundar Rajan}
\authorblockA{Dept. of ECE, IISc, Bangalore 560012, India, Email:{$\lbrace$tmvijay, bsrajan$\rbrace$}@ece.iisc.ernet.in
}
}

\maketitle
% \thispagestyle{empty}	
%%%%%%%%
\begin{abstract}
The design of modulation schemes for the physical layer network-coded two-way MIMO relaying scenario is considered, with $n_R$ antennas at the relay R, $n_A$ and $n_B$ antennas respectively at the end nodes A and B. We consider the denoise-and-forward (DNF) protocol which employs two phases: Multiple access (MA) phase and Broadcast (BC) phase. It is known for the network-coded SISO two-way relaying that adaptively changing the networking coding map used at the relay, also known as the denoising map, according to the channel conditions greatly reduces the impact of multiple access interference which occurs at the relay during the MA phase and all these network coding maps should satisfy a requirement called the {\it exclusive law}. The network coding maps which satisfy exclusive law can be viewed equivalently as Latin Rectangles. In this paper, it is shown that for MIMO two-way relaying, deep fade occurs at the relay when the row space of the channel fade coefficient matrix is a subspace of a finite number of vector subspaces of $\mathbb{C}^{n_A+n_B}$ which are referred to as the singular fade subspaces. It is shown that proper choice of network coding map can remove most of the singular fade subspaces, referred to as the removable singular fade subspaces. All these network coding maps are obtainable by the completion of partially filled Latin Rectangles. For $2^{\lambda}$-PSK signal set, the number of removable and non-removable singular fade subspaces are obtained analytically and it is shown that the number of non-removable singular fade subspaces is a small fraction of the total number of  singular fade subspaces. The Latin Rectangles for the case when the end nodes use different number of antennas are shown to be obtainable from the Latin Squares for the case when they use the same number of antennas, irrespective of the value of $n_R$. For $2^{\lambda}$-PSK signal set, the singular fade subspaces which are removed by the conventional XOR network code are identified. Also, using the notions of isotopic and transposed Latin Squares, the network coding maps which remove all the removable singular singular fade subspaces are shown to be obtainable from a small set of Latin Squares.
\end{abstract}

\section{Background and Preliminaries}
\begin{figure}[htbp]
\centering
\subfigure[MA Phase]{
\includegraphics[totalheight=1.5in,width=3in]{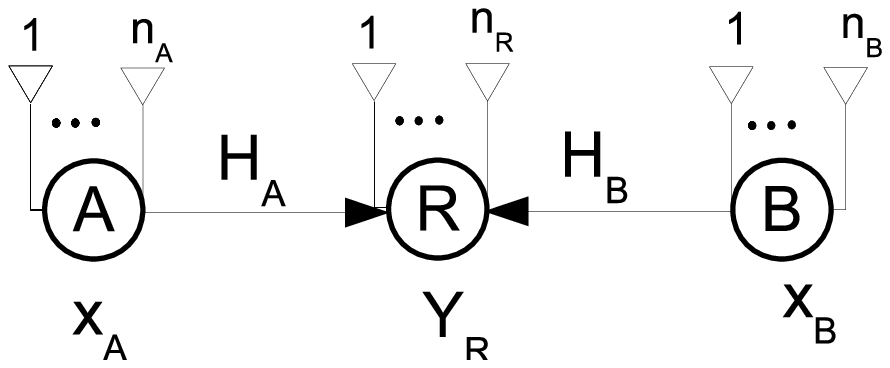}
\label{fig:phase1}
}
\subfigure[BC Phase]{
\includegraphics[totalheight=1.5in,width=3in]{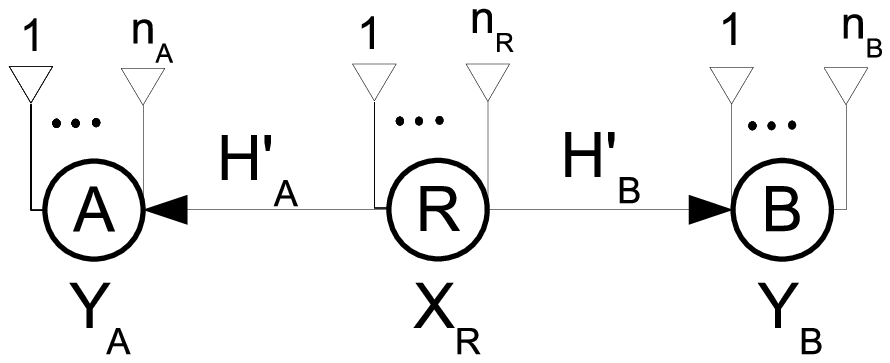}
\label{fig:phase2}
}
\caption{The Two Way Relay Channel}
\label{relay_channel}
\end{figure}

We consider the two-way wireless relaying scenario shown in Fig.\ref{relay_channel} with multiple antennas at the nodes, where data transfer takes place between the nodes A and B with the help of the relay R. It is assumed that all the three nodes operate in half-duplex mode, i.e., they cannot transmit and receive simultaneously in the same frequency band. We consider the denoise-and-forward (DNF) protocol originally introduced in \cite{PoYo_DNF}, which consists of the following two phases: the \textit{multiple access} (MA) phase, during which A and B simultaneously transmit to R and the \textit{broadcast} (BC) phase during which R transmits to A and B. Network coding map, which is also referred to as the denoising map, is chosen at R in such a way that A (B) can decode the messages of B (A), given that A (B) knows its own messages. 

%%%%%%%%%%%%%%%%%%%%%%%%%%%%%%%%%%%
%\begin{figure}[htbp]
%\centering
%\subfigure[MA Phase]{
%\includegraphics[totalheight=1in,width=2in]{2way_relay_MAC}
%\label{fig:phase1}
%}
%
%\subfigure[BC Phase]{
%\includegraphics[totalheight=1in,width=2in]{2way_relay_BC}
%\label{fig:phase2}
%}
%\caption{The Two Way Relay Channel}
%\label{relay_channel}
%\end{figure}
%\vspace{-0.1 cm}
%%%%%%%%%%%%%%%%%%%%%%%%%%%%%%%%
%Most of the background and introductory material in this section can also be found in  \cite{VNR} also. 
%%%%%%%%%%%%%%%%%%%%%%%%%%%%%% 
%\subsection{Background}
 
 The concept of physical layer network coding has attracted a lot of attention in recent times. The idea of physical layer network coding for the two way relay channel was first introduced in \cite{ZLL}, where the multiple access interference occurring at the relay was exploited so that the communication between the end nodes can be done using a two phase protocol. Information theoretic studies for the physical layer network coding scenario were reported in \cite{KMT},\cite{PoY}. A differential modulation scheme with analog network coding for two-way relaying was proposed in \cite{LiYoAnBiAt}. The design principles governing the choice of modulation schemes to be used at the nodes for uncoded transmission were studied in \cite{APT1}. An extension for the case when the nodes use convolutional codes was done in \cite{APT2}. A multi-level coding scheme for the two-way relaying scenario was proposed in \cite{HeN}. Power allocation strategies and lattice based coding schemes for two-way relaying were proposed in \cite{WiNa}.

It was observed in \cite{APT1} that for uncoded transmission, the network coding map used at the relay needs to be changed adaptively according to the channel fade coefficient, in order to minimize the impact of the Multiple Access Interference (MAI). For the single antenna two-way relaying scenario, a computer search algorithm called the \textit{Closest-Neighbour Clustering} (CNC) algorithm was proposed in \cite{APT1} to obtain the adaptive network coding maps resulting in the best distance profile at R. An extension to MIMO two-way relaying using the CNC algorithm was made in \cite{Ak_MIMO}. With a single antenna at the nodes, the MAI becomes severe for channel fade states referred to as the singular fade states \cite{NVR}. An alternative procedure to obtain the network coding maps, based on the removal of singular fade states using Latin Squares was proposed in \cite{NVR}. A quantization of the set of all possible channel realizations based on the network code used was obtained analytically in \cite{VNR}.

In this paper, it is shown that for the MIMO two-way relaying scenario, the MAI becomes severe when the row space of the channel fade coefficient matrix is a subspace of a finite number of vector subspaces of $\mathbb{C}^{n_A+n_B}$ referred to as the singular fade subspaces. The notion of singular fade subspaces subsumes in it as a special case the notion of singular fade states introduced in \cite{NVR} for the single antenna two-way relaying scenario. It is shown that the network coding maps totally avoiding the removable singular fade subspaces can be obtained by the completion of partially filled Latin Rectangles.

\textbf{\textit{Notations}:}
  Let $\mathcal{CN}(0,\sigma^2I_{n})$ denote the circularly symmetric complex Gaussian random vector of length $n$ with covariance matrix $\sigma ^2I_{n}$, where $I_n$ is the identity matrix of order $n$. The binary field is denoted by $\mathbb{F}_2.$ All the vector spaces and vector subspaces considered in this paper are over the complex field $\mathbb{C}.$ Let $\text{span}(c_1,c_2, \dotso c_L)$ denote the vector space over $\mathbb{C}$ spanned by the complex vectors $c_1,c_2, \dotso c_L \in \mathbb{C}^n.$ For a matrix $A,$ $A^T$ denotes
its transpose. For a vector subspace $V$ of $\mathbb{C}^{n},$ $V^{\perp}$ denotes the vector subspace $\lbrace x: x^T v =0 \: \forall v \in V\rbrace.$ For a matrix $H \in \mathbb{C}^{m \times n},$ $\mathcal{R}(H)$ denotes the row space of $H.$ The all zero vector of length $n$ is denoted by $0_{n}.$ For a vector $x$ of length $n,$ $x_i, 1 \leq i \leq n$ denotes the $i^{th}$ component of $x.$ For vector subspaces $V_1$ and $V_2$ of $\mathbb{C}^{n},$ $V_1  \preceq  V_2$ means that $V_1$ is a subspace of $V_2$ and $V_1 \npreceq V_2$ means that $V_1$ is not a subspace of $V_2.$ By $n_A \times n_B$ system, we refer to the two-way relaying system with $n_A$ and $n_B$ antennas at the nodes A and B respectively, with no restriction placed on the number of antennas at the relay $n_R$. For a vector $x,$ $x_{[i : j]}, 1 \leq i \leq j \leq n$ denotes the vector obtained by taking only the $i^{th}$ to $j^{th}$ components of $x.$ %For vector $x$, $\sqsubset x \sqsupset$ denotes the vector whose components are the absolute values of the components of $x.$

%The ratio of the channel fade states corresponding to the A-R and B-R links determined the choice of the network coding map \cite{NVR_arxiv}.  

%In this paper, it is shown that a finite number of network coding maps, obtained by completion of partially filled Latin Squares, can effectively mitigate the effect of multiple access interference. Having obtained the network coding maps, the set of all possible channel realizations is quantized into a finite number of regions, with a specific network coding map giving the best performance in a particular region. In \cite{APT1}, a quantization was obtained using exhaustive computer search for 4-PSK signal set.  In this paper, we obtain a quantization analytically for any $M$-PSK signal set, for $M$ any power of 2. 

%%%%%%%%%%%%%%%%%%%%%%%%%%%%%%%%%%%%%%%%%%%%%%%%%%
\subsection{Signal Model}

Consider the MIMO two-way relaying system with $n_A,$ $n_R$ and $n_B$ antennas at the nodes A, R and B respectively as shown in Fig. \ref{relay_channel}.   
%Throughout the paper, the size of the constellation $M$ used in each one of the antennas at the end nodes is assumed to be a power of 2, unless mentioned explicitly as arbitrary $M$. 
It is assumed that the complex numbers transmitted in each one of the antennas at the end nodes belong to the signal set $\mathcal{S}$ of size $M=2^{\lambda}, $ where $\lambda$ is an integer. Assume that A (B) wants to transmit a $\lambda n_A$-bit ($\lambda n_B$-bit) binary tuple to B (A). At node A (B), the $\lambda n_A$ ($\lambda n_B$) bits are spatially multiplexed into $n_A$ ($n_B$) streams with each one of the streams consisting of $\lambda$ bits. The $\lambda$ bits in each one of the streams are mapped on to the signal set $\mathcal{S}$ using the map $\mu: \mathbb{F}_2^\lambda \rightarrow \mathcal{S}$ and are transmitted. %Throughout, the points in the $M$-PSK signal set are assumed to be of the form $e^{j (2k+1) \pi/M},0 \leq k \leq M-1.$ An element $e^{j\frac{(2k+1)\pi}{M}} \in \mathcal{S}$ will be simply denoted by $k$ throughout and all the results presented in the paper hold with no restriction on the mapping $\mu.$  For all illustrative purposes we use the map binary-to-decimal conversion map.

\subsubsection*{Multiple Access (MA) phase}
Throughout, we assume a block fading scenario with the Channel State Information (CSI) available only at the receivers. Let $x_A= [\mu(s_{A_1}),\mu(s_{A_2}),...,\mu(s_{A_{n_A}})]^T \in \mathcal{S}^{n_A}$, $x_B= [\mu(s_{B_1}),\mu(s_{B_2}),...,\mu(s_{B_{n_B}})]^T \in \mathcal{S}^{n_B}$ denote the complex vectors transmitted by A and B respectively, where $s_{A_1},s_{A_2},...,s_{A_{n_A}},s_{B_1},s_{B_2},...,s_{B_{n_B}} \in \mathbb{F}_2^\lambda$. The received signal at $R$ is given by,
\begin{align}
\nonumber
Y_R=H_{A} x_A + H_{B} x_B +Z_R,
\end{align}
where $H_A$ of size $n_R \times n_A$ and $H_B$ of size $n_R \times n_B$ are the channel matrices associated with the A-R and B-R links respectively. The additive noise vector $Z_R$ is assumed to be $\mathcal{CN}(0,\sigma^2 I_{n_R})$.
%We assume a block fading scenario, with the ratio $ H_{B}/H_{A}$ denoted as $z=\gamma e^{j \theta}$, where $\gamma \in \mathbb{R}^+$ and $-\pi \leq \theta < \pi,$ is referred as the {\it fade state} and for simplicity, also denoted by $(\gamma, \theta).$ Also, it is assumed that $z$ is distributed according to a continuous probability distribution. Throughout the paper, all the distances are assumed to be normalized with respect to $\vert H_A \vert$. 

% 
 Let $\mathcal{S}_{R}(H_A,H_B) \subset \mathbb{C}^{n_R}$ denote the effective constellation seen at the relay during the MA phase, i.e., 
\begin{align} 
\nonumber
\mathcal{S}_{R}(H_A,H_B) =\left\lbrace H_A x_A+ H_B x_B \vert x_A \in \mathcal{S}^{n_A}, x_B \in \mathcal{S}^{n_B}\right \rbrace.
\end{align}
 
 Let $d_{min}(H_A,H_B)$ denote the minimum distance between the points in the effective constellation $\mathcal{S}_{R}(H_A,H_B)$, i.e.,

{\footnotesize
\begin{align}
\label{eqn_dmin} 
d_{min}(H_A,H_B)=\hspace{-0.5 cm}\min_{\substack {{(x_A,x_B),(x'_A,x'_B) \in \mathcal{S}^{n_A+n_B}} \\ {(x_A,x_B) \neq (x'_A,x'_B)}}}\hspace{-1 cm}\parallel H_A\left(x_A-x'_A\right)+H_B \left(x_B-x'_B\right)\parallel.
\end{align}
}
 From \eqref{eqn_dmin}, it is clear that there exists values of $H_A,H_B$ for which $d_{min}(H_A,H_B)=0$. 
 
\begin{figure}[htbp]
\vspace{-.4 cm}
\centering
\includegraphics[totalheight=2.5in,width=3.6in]{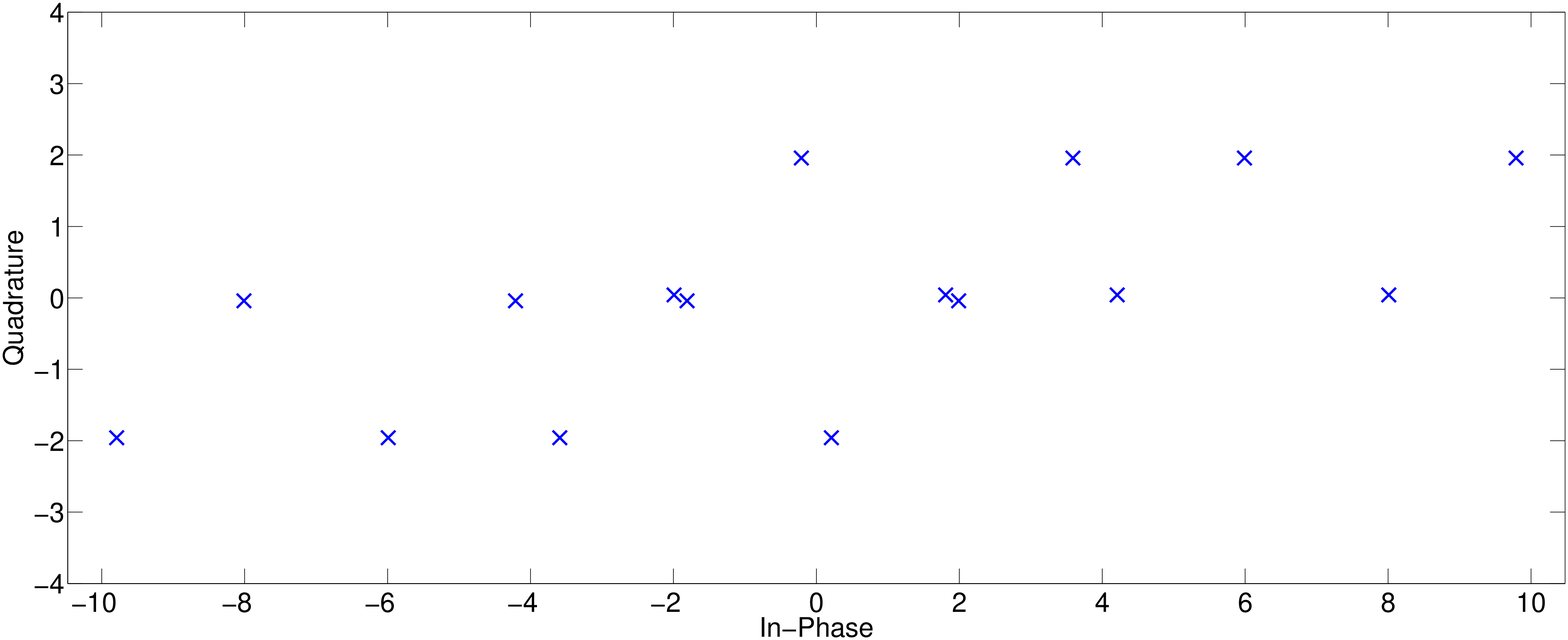}
\vspace{-1 cm}
\caption{Diagram showing the effective constellation at the relay for $H_R \approx [1+j \quad 2\quad 3 \quad 4+j]^T$}
\label{fig:dist_shortening}
\end{figure}

\begin{example}
Consider the case when $n_A=n_B=2$ and $n_R=1$ and BPSK signal set $\lbrace \pm 1 \rbrace$ is used at the end nodes. When $[H_A \: H_B] \approx [1+j \quad 2\quad 3 \quad 4+j]^T,$ the effective constellation $\mathcal{S}_R$ is as shown in Fig. \ref{fig:dist_shortening}. It is clear from Fig. \ref{fig:dist_shortening} that for $[H_A \: H_B] = [1+j \quad 2\quad 3 \quad 4+j]^T,$ $d_{min}(H_A,H_B)=0.$
\end{example}

Let $\Delta \mathcal{S}$ denote the difference constellation of the signal set $\mathcal{S},$ i.e., $\Delta\mathcal{S}=\lbrace s_i-s'_i \vert  s_i, s'_i \in \mathcal{S}\rbrace.$

 Let us define $\Delta x=[(x_A-x'_A)^T \:(x_B-x'_B)^T]^T \in \Delta \mathcal{S}^{n_A+n_B}$, where $\Delta {x_{A}}=x_A-x'_A \in \Delta \mathcal{S}^{n_A}$ and $\Delta {x_{B}}=x_B-x'_B \in \Delta \mathcal{S}^{n_B}$, and $H_R=[H_A \: H_B].$ Also let $H_R=[h_1 \: h_2 \:  ... h_{n_R}]^T$, where the row vector ${h^T_k}, 1 \leq k \leq n_R$ of length $n_A+n_B$ is the $k^\mathrm{th}$ row of $H_R.$ From \eqref{eqn_dmin}, it follows that,
 
 {\begin{align}
\nonumber
d_{min}^2(H_A,H_B)&=\min_{{{\Delta x \in \Delta \mathcal{S}^{n_A+n_B}} , {\Delta x \neq  0_{n_A+n_B}}}}\parallel H_R \Delta x\parallel ^2.\\
\label{eqn_dmin1}
&=\min_{ {{\Delta x \in \Delta \mathcal{S}^{n_A+n_B}} ,{\Delta x \neq  0_{n_A+n_B}}}} \sum_{k=1}^{n_R}\vert h^T_k \Delta x\vert ^2.
\end{align}
}

From \eqref{eqn_dmin1} it follows that $d_{min}^2(H_A,H_B)=0$ if $H_R$ is such that $h^T_k \Delta x =0, \forall 1 \leq k \leq n_R$, for some $\Delta x \in \Delta \mathcal{S} ^{n_A+n_B}.$ Equivalently, for $d_{min}^2(H_A,H_B)=0$, the vectors  $h_k, 1 \leq k \leq n_R$ should belong to the vector subspace $[\mathrm{span}(\Delta x)]^\perp$ for some $\Delta x \in \Delta \mathcal{S} ^{n_A+n_B}.$ In other words, the row space of the matrix $H_R$ should be a subspace of the vector subspace $[\mathrm{span}(\Delta x)]^\perp,$ for some $\Delta x \in \Delta S ^{n_A+n_B},$ for $d_{min}^2(H_A,H_B)$ to become zero. 

The channel fade coefficient matrix $H_R$ is said to be a \textit{deep fade matrix}, if $d_{min}(H_A,H_B)=0.$ The row space of the deep fade matrices are referred to as the \textit{deep fade spaces}. The deep fade spaces are subspaces of the vector subspaces $[\mathrm{span}(\Delta x)]^\perp, \Delta x \in \Delta S ^{n_A+n_B},$ which are referred to as the \textit{singular fade subspaces}. Let $\mathcal{F}$ denote the set of all singular fade subspaces, i.e., $\mathcal{F}=\lbrace [\mathrm{span}(\Delta x)]^\perp  :  \Delta x \in \Delta S ^{n_A+n_B} \rbrace.$

\begin{note}
Note that singular fade subspaces depend only on the set $\lbrace \Delta x : \Delta x \in \Delta \mathcal{S}^{n_A+n_B}\rbrace,$ i.e., they are independent of $n_R.$\\
\end{note} 
\textit{SISO two-way relaying as a special case:}

Consider the case when $n_A=n_B=n_R=1.$ The row space of $H_R,$ $\mathcal{R}(H_R)=k [H_A \: H_B]^T = k' [1 \quad \frac{H_B}{H_A}]^T, \mathrm{ where} \: k, k' \in \mathbb{C}.$ For this case, $\mathrm{span}(\Delta x)=c[\Delta x_A \: \Delta x_B]^T,$ where $c \in \mathbb{C}$ and $\Delta x_A , \Delta x_B \in \Delta \mathcal{S}.$ Hence, the singular fade subspaces $[\mathrm{span}(\Delta x)]^\perp, \Delta x \in \Delta \mathcal{S}^2$ are of the form $c'[1 \quad \frac{-\Delta x_A}{\Delta x_B}]^T, \: \mathrm{where} \: c' \in \mathbb{C} .$ Note that for this case, $\mathcal{R}(H_R)$ and $[\mathrm{span}(\Delta x)]^\perp$ are both one dimensional subspaces of the two dimensional vector space $\mathbb{C}^2$ over $\mathbb{C}$. Hence for $\mathcal{R}(H_R)$ to be a subspace of $[\mathrm{span}(\Delta x)]^\perp,$ it is necessary that $\mathcal{R}(H_R)=[\mathrm{span}(\Delta x)]^\perp.$ It can be verified that $\mathcal{R}(H_R)=[\mathrm{span}(\Delta x)]^\perp$ if and only if $\frac{H_B}{H_A}=\frac{-\Delta x_A}{\Delta x_B}.$ Hence, the effect of the MAI is totally captured by the ratio of the channel fade coefficients $\frac{H_B}{H_A},$ consistent with the results in \cite{APT1}, \cite{NVR}. Also, when the ratio $\frac{H_B}{H_A}$ referred to as the fade state in \cite{NVR} becomes equal to $\frac{-\Delta x_A}{\Delta x_B}$ for some $\Delta x_A , \Delta x_B \in \Delta \mathcal{S},$ the minimum distance of the effective constellation at the relay becomes zero. The complex numbers $\frac{-\Delta x_A}{\Delta x_B}, \:\text{where}\: \Delta x_A, \Delta x_B \in \Delta \mathcal{S},$ were referred to as the singular fade states in \cite{NVR}.

 From above, it is clear that the notion of singular fade subspaces subsumes in it as a special case the notion of singular fade states used for two-way relaying with single antenna at the nodes.
 
\begin{example}
Consider the $2 \times 2$ system with BPSK signal set used at nodes A and B, with the bit $0$ mapped onto $+1$ and the bit $1$ mapped onto $-1$. The difference constellation of the BPSK signal set $\Delta \mathcal{S}=\lbrace -2, 0,2\rbrace.$ The vector subspace $f=[\mathrm{span}([2 \; 2 \; 2 \; 2]^T)]^\perp=[\mathrm{span}([-2 \: -2 \:-2 \:-2]^T)]^\perp$ is a singular fade subspace for this case.
\end{example}
\subsubsection*{Broadcast (BC) phase}

Let $(\hat{x}_A,\hat{x}_B) \in \mathcal{S}^{n_A+n_B}$ denote the Maximum Likelihood (ML) estimate of $({x}_A,{x}_B)$ at R based on the received complex vector $Y_{R}$, i.e.,
\begin{align}
(\hat{x}_A,\hat{x}_B)=\arg\min_{({x}'_A,{x}'_B) \in \mathcal{S}^{n_A+n_B}} \vert Y_R-H_{A}{x}'_A-H_{B}{x}'_B\vert.
\end{align}

Depending on $H_{R}$, R chooses a many-to-one map $\mathcal{M}^{H_R}:\mathcal{S}^{n_A+n_B} \rightarrow \mathcal{S}'$, where $\mathcal{S}' \subset \mathbb{C}^{n_R}$ is the signal set (of size between $\max\left \lbrace M^{n_A}, M^{n_B} \right \rbrace$ and $M^{n_A+n_B}$) used by R during the $BC$ phase. Note that $\vert \mathcal{S}'\vert$ should be at least $\max\left \lbrace M^{n_A}, M^{n_B} \right \rbrace$, to transmit $\max\lbrace \lambda n_A, \lambda n_B \rbrace$ information bits. The elements in $\mathcal{S}^{n_A+n_B}$ which are mapped on to the same complex vector in $\mathcal{S}'$ by the map $\mathcal{M}^{H_R}$ are said to form a cluster. Let $\lbrace \mathcal{L}_0, \mathcal{L}_2,...,\mathcal{L}_{t-1}\rbrace$ denote the set of all such clusters. The formation of clusters for $H_R$ is called clustering, and is denoted by $\mathcal{C}^{H_R}.$ For simplicity, in the rest of the paper, the cluster $\mathcal{L}_k$ is denoted by the subscript $k,$ where $0 \leq k \leq t-1.$

\begin{example}
For the $2 \times 2$ system with BPSK signal set,

\begin{equation}
\label{clustering1}
\begin{Bmatrix}
\begin{Bmatrix}
[0 \; 0 \; 0 \;0]^T, [0 \; 1 \; 0 \; 1]^T, [1 \; 0 \; 1 \; 0]^T, [1 \; 1 \; 1 \; 1]^T 
\end{Bmatrix}
\\
\begin{Bmatrix}
[0 \; 0 \; 0 \;1]^T, [0 \; 1 \; 0 \; 0]^T, [1 \; 0 \; 1 \; 1]^T, [1 \; 1 \; 1 \; 0]^T \\
\end{Bmatrix}
\\
\begin{Bmatrix}
[0 \; 0 \; 1 \;0]^T, [0 \; 1 \; 1 \; 1]^T, [1 \; 0 \; 0 \; 0]^T, [1 \; 1 \; 0 \; 1]^T \\
\end{Bmatrix}
\\
\begin{Bmatrix}
[0 \; 0 \; 1 \;1]^T, [0 \; 1 \; 1 \; 0]^T, [1 \; 0 \; 0 \; 1]^T, [1 \; 1 \; 0 \; 0]^T \\
\end{Bmatrix}
\end{Bmatrix},
\end{equation}
represents a clustering with four clusters. For two decoded pairs $(x_A,x_B)$ and $(x'_A,x'_B),$ the relay transmits the same vector from the signal set $\mathcal{S'}$ if $[x_A \: x_B]$ and $[x'_A \: x'_B]$ belong to the same cluster. For example, if R uses the clustering given in \eqref{clustering1}, the vector transmitted during the BC phase will be the same, if the decoded pair during MA phase is  $([0 \; 0]^T,[0 \;0]^T)$ or $([0 \; 1]^T,[0 \;1]^T),$ since $[0 \; 0 \; 0 \;0]^T$ and $[0 \; 1 \; 0 \; 1]^T$ belong to the same cluster.
\end{example}

The received signals at A and B during the BC phase are respectively given by,
\begin{align}
Y_A=H'_{A} X_R + Z_A,\;Y_B=H'_{B} X_R + Z_B,
\end{align}
where $X_R=\mathcal{M}^{H_R}(\hat{x}_A,\hat{x}_B) \in \mathcal{S'}$ is the complex vector transmitted by R. The fading matrices of size $n_A \times n_R$ and $n_B \times n_R$ corresponding to the R-A and R-B links are denoted by $H'_{A}$ and $H'_{B}$ respectively and the additive noises $Z_A$ and $Z_B$ are $\mathcal{CN}(0,\sigma ^2$).

In order to ensure that A (B) is able to decode B's (A's) messages, the clustering $\mathcal{C}^{H_R}$ should satisfy the exclusive law \cite{APT1}, i.e.,

{\footnotesize
\begin{align}
\left.
\begin{array}{ll}
\nonumber
\mathcal{M}^{H_R}(x_A,x_B) \neq \mathcal{M}^{H_R}(x'_A,x_B), \forall x_A \neq x'_A \in \mathcal{S}^{n_A}, x_B \in  \mathcal{S}^{n_B},\\
\nonumber
\mathcal{M}^{H_R}(x_A,x_B) \neq \mathcal{M}^{H_R}(x_A,x'_B), \forall x_B \neq x'_B \in \mathcal{S}^{n_B}, x_A \in \mathcal{S}^{n_A}.
\end {array}
\right\} \\
\label{ex_law}
\end{align}
\vspace{-.3 cm}
}

\begin{definition}
The cluster distance between a pair of clusters $\mathcal{L}_i$ and $\mathcal{L}_j$ is the minimum among all the distances calculated between the points $H_A x_A+ H_B x_B$ and $H_A x'_A+ H_B x'_B \in \mathcal{S}_R(H_A,H_B)$, where $(x_A,x_B) \in \mathcal{L}_i$ and $(x'_A,x'_B) \in \mathcal{L}_j$. The \textit{minimum cluster distance} of the clustering $\mathcal{C}^{H_R}$ is the minimum among all the cluster distances, i.e.,

{\footnotesize
\begin{align}
\nonumber
d_{min}(\mathcal{C}^{H_R})=\hspace{-0.8 cm}\min_{\substack {{(x_A,x_B),(x'_A,x'_B) \in \mathcal{S}^{n_A+n_B},} \\ {\mathcal{M}^{H_R}(x_A,x_B) \neq \mathcal{M}^{H_R}(x'_A,x'_B)}}}\hspace{-0.8 cm}  \parallel H_A \left( x_A-x'_A\right)+ H_B \left(x_B-x'_B\right) \parallel.
\end{align}
}

\end{definition}
%%Let $\mathcal{H}^\mathcal{C}=\lbrace \gamma e^{j\theta} \in \mathbb{C} \vert d_{min}^\mathcal{C}=0 \rbrace$. The elements of $\mathcal{H}^\mathcal{C}$ are called the singularity points of the clustering $\mathcal{C}$.

The minimum cluster distance determines the performance during the MA phase of relaying. The performance during the BC phase is determined by the minimum distance of the signal set $\mathcal{S}'$. Throughout, we restrict ourselves to optimizing the performance during the MA phase. For values of $H_R$ such that $\vert h_k^T\Delta x\vert$ is small, $\forall 1 \leq k \leq n_R,$ for some $\Delta x \in \Delta \mathcal{S}^{n_A+n_B},$ $d_{min}(H_R)$ is greatly reduced, a phenomenon referred as {\it distance shortening}. 
To avoid distance shortening, for every removable singular fade subspace, a clustering needs to be chosen such that the minimum cluster distance at every $H_R$ whose rows belong to that singular fade subspace is non-zero.  
%For values of $H_R$ such that the perpendicular distances between each one of the the rows of $H_R$ and one of the singular fade  spaces are small, $d_{min}(H_R)$ is greatly reduced, a phenomenon referred as {\it distance shortening}. 

For a singular fade subspace $f \in \mathcal{F}$, let $d_{min}(\mathcal{C}^{f},H_R)$ be defined as,

{\footnotesize
\begin{align}
\nonumber
d_{min}(\mathcal{C}^{f},H_R)=\hspace{-0.8 cm}\min_{\substack {{(x_A,x_B),(x'_A,x'_B) \in \mathcal{S}^{n_A+n_B},} \\ {\mathcal{M}^{f}(x_A,x_B) \neq \mathcal{M}^{f}(x'_A,x'_B)}}}\hspace{-0.8 cm}\vert H_A\left( x_A-x'_A\right)+H_B \left(x_B-x'_B\right)\vert,
\end{align}
}where $\mathcal{M}^{f}$ is the many-one map associated with the clustering $\mathcal{C}^{f}.$

 A clustering $\mathcal{C}^{ f }$ is said to remove a singular fade subspace $ f \in \mathcal{F}$, if the minimum cluster distance $d_{min}(\mathcal{C}^{f},H_R)$ is greater than zero, for every $H_R$ such that $\mathcal{R}(H_R) \preceq f$. If there are more than one clusterings which remove a singular fade subspace $f$, choose any one of them. 
 
 It is important to note that certain singular fade subspaces cannot be removed. These are precisely the singular fade subspaces which are of the form $[\mathrm{span}(\Delta x)]^\perp,$ for which 
 \begin{align*} 
&\Delta x= [ 0_{n_A} \; \Delta x_{B}], \Delta x_{B} \in \Delta \mathcal{S}^{n_B} \textrm{\; or \;} \\
&\Delta x=[\Delta x_{A} \; 0_{n_B}], \Delta x_{A} \in \Delta \mathcal{S}^{n_A},
\end{align*} 
and are referred to as the non-removable singular fade subspaces. The reason for this is as follows: The pair $(x_A,x_B)$ and $(x_A,x'_B)$ result in $\Delta x= [ 0_{n_A} \; \Delta x_{B}].$ But $(x_A,x_B)$ and $(x_A,x'_B)$ cannot be placed in the same cluster since exclusive law given in \eqref{ex_law} will be violated. Note that $0_{n_A+n_B}$ is also a non-removable singular fade subspace, referred to as the trivial non-removable singular fade subspace.

\begin{note}
The non-trivial non-removable singular fade subspaces for the SISO two-way relaying scenario are of the form $[\text{span}([\Delta x \quad 0]^T)]^\perp= \text{span}([0 \; 1]^T)$ and $[\text{span}([0 \quad \Delta x]^T)]^\perp= \text{span}([1 \; 0]^T),$ where $\Delta x \in \Delta \mathcal{S} \setminus \lbrace 0 \rbrace.$ Irrespective of the signal set used at A and B, the non-trivial non-removable singular fade spaces are only two in number, since the vector subspaces $[\text{span}([\Delta x \quad 0]^T)]^\perp$ are the same for all $\Delta x \in \Delta \mathcal{S} \setminus \lbrace 0 \rbrace.$ In terms of the notion of singular fade states \cite{NVR}, from the earlier discussion on SISO two-way relaying as a special case of MIMO two-way relaying, it follows that the above two singular fade subspaces correspond to the singular fade states infinity and zero respectively. Alternatively, the fact that these singular fade states are non-removable has been stated in \cite{APT1} as follows: \textit{the distance shortening at $H_B/H_A \approx 0$ (or $H_A/H_B \approx 0$) is inevitable}. For SISO two-way relaying, the number of non-trivial non-removable singular fade subspaces remains two irrespective of the size of the signal set used at A and B. In contrast, it will be seen in Section II that for MIMO two-way relaying, the number of non-trivial non-removable singular fade subspaces increases with increasing size of the signal set used at A and B.
\end{note}
 %For a singular fade subspace $f \in \mathcal{F}$, let $\mathcal{C}_{\lbrace f\rbrace}$ denote a clustering which removes the singular fade state $h$ 

Let $\mathcal{C}_{\mathcal{F}}=\left\lbrace \mathcal{C}^{ f} : f \in \mathcal{F} \right\rbrace$ denote the set of all clusterings, which remove a removable singular fade subspace. 

\begin{note}
For every removable singular fade subspace, the set $\mathcal{C}_{\mathcal{F}}$ contains exactly one clustering which removes that singular fade subspace. The clusterings which belong to the set $\mathcal{C}_{\mathcal{F}}$ need not be distinct, since a single clustering can remove more than one singular fade subspace, as shown in the following example.
\end{note}

\begin{example}
Consider the $2 \times 2$ system with BPSK signal set used at nodes A and B, with the bit $0$ mapped onto $+1$ and the bit $1$ mapped onto $-1$. The difference constellation of the BPSK signal set is $\Delta \mathcal{S}=\lbrace -2, 0,2\rbrace.$ Consider the singular fade subspace, $f=[\mathrm{span}([2 \; 2 \; 2 \; 2]^T)]^\perp=[\mathrm{span}([-2 \: -2 \:-2 \:-2]^T)]^\perp.$ The binary vectors $[s_{A_1} \; s_{A_2} \; s_{B_1} \; s_{B_2}]^T, [s'_{A_1} \; s'_{A_2} \; s'_{B_1} \; s'_{B_2}]^T \in \mathbb{F}_2^4$ which result in the singular fade subspace $f$ are $[0 \; 0 \; 0 \; 0]^T, [1 \; 1 \; 1 \;1]^T.$ The above two vectors need to be placed in the same cluster by the clustering which removes the singular fade subspace $f.$ The clustering given in \eqref{clustering1} in Example 3 removes the singular fade subspace $f.$
Since $[0 \; 0 \; 1 \;1]^T$ and $[1 \; 1 \; 0 \; 0]^T$ are in the same cluster, the clustering given in \eqref{clustering1} removes the singular fade subspace $f'=[\mathrm{span}([2 \quad 2 \quad -2 \quad -2]^T)]^\perp=[\mathrm{span}([-2 \quad -2 \quad 2 \quad 2]^T)]^\perp$ as well.

%if the fade state is $(\gamma=1,\theta=0)$ the distance between the pairs $(0,1)(1,0)$ is zero as in Fig. \ref{fig:BPSK}(a). 

%Similarly, when $(\gamma=1,\theta=\pi),$ the distance between the pairs $(0,0)(1,1)$ is zero as in Fig. \ref{fig:BPSK}(b). The following clustering removes both the singular fade states $(\gamma=1,\theta=0)$ and $(\gamma=1,\theta=\pi)$:
%$$\{\{(0,1)(1,0)\},\{(1,1)(0,0)\}\}.$$
%%The minimum cluster distance is non zero for this clustering.
\end{example}
%

%Let $d_{min}(\mathcal{C}^{f},H_R)$ be defined as,
%
%{\footnotesize
%\begin{align}
%\nonumber
%d_{min}(\mathcal{C}^{f},H_R)=\hspace{-0.8 cm}\min_{\substack {{(x_A,x_B),(x'_A,x'_B) \in \mathcal{S}^{n_A+n_B},} \\ {\mathcal{M}^{\lbrace h\rbrace}(x_A,x_B) \neq \mathcal{M}^{\lbrace h \rbrace}(x'_A,x'_B)}}}\hspace{-0.8 cm}\vert H_A \left( x_A-x'_A\right)+ H_B \left(x_B-x'_B\right)\vert.
%\end{align}
%}
%
%The quantity $d_{min}({\mathcal{C}^{\lbrace h\rbrace}},\gamma,'\theta')$ is referred to as  the minimum cluster distance of the clustering $\mathcal{C}^{\lbrace h\rbrace}$ evaluated at $H_R.$
%
 In general, the row space of $H_R$ need not be a subspace of a singular fade subspace. In such a scenario, among all the clusterings which remove the singular fade subspaces, the one which maximizes the minimum cluster distance is chosen. In other words, for $\mathcal{R}({H_R}) \npreceq \mathcal{F}$, the clustering $\mathcal{C}^{H_R}$ is chosen to be $\mathcal{C}^{f}$, which satisfies $d_{min}(\mathcal{C}^{f},H_R) \geq d_{min}({\mathcal{C}^{f'}},H_R), \forall f \neq f' \in \mathcal{F}$. Since the clusterings which remove the singular fade subspaces are known to all the three nodes and are finite in number, the clustering used for a particular realization of the channel fade coefficients can be indicated by R to A and B using overhead bits.

In \cite{APT1}, a computer search algorithm called the Closest-Neighbour Clustering (CNC) algorithm was proposed for two-way relaying with single antenna at the nodes, which was used to obtain the network coding map that results in the best distance profile. The CNC algorithm was extended to the multiple antenna scenario in \cite{Ak_MIMO}. The algorithm is run for a given $H_R.$ The total number of network coding maps which would result is known only after the algorithm is run for all possible realizations of  $H_R$ which is uncountably infinite. Hence, the number of overhead bits required is not known beforehand.

In contrast, the scheme proposed in this paper is based on the removal of singular fade subspaces. Since the number of singular fade subspaces is finite, the number of overhead bits required is upper bounded by the number of singular fade subspaces, which is known beforehand. In other words, the total number of network coding maps required is known exactly, which determines the number of overhead bits required. It is shown in Section III that the problem of obtaining clusterings which remove all the singular fade subspaces reduces to completing a finite number of partially filled Latin Rectangles, which totally avoids the problem of performing exhaustive search for an uncountably infinite number of values. 

The contributions and organization of this paper are as follows.

\begin{itemize}
\item
The structure and the exact number of non-removable and removable singular fade subspaces for $M-PSK$ signal set ($M$ any power of 2) are obtained analytically (Section II). It is shown that the fraction of the number of non-removable singular fade subspaces to the total number of singular fade subspaces tends to zero for large values of $M$ (Section II). 

\item
It is shown that the requirement of satisfying the exclusive law is same as the clustering being represented by a partially filled Latin Rectangle (PFLR) and can be used to get the clustering which removes singular fade subspaces, by completing the PFLR (Section III A).
\item
It is shown that the Latin Rectangles which remove the singular fade subspaces for the case when end nodes have unequal number of antennas, i.e., $n_A \neq n_B$ can be obtained from the Latin Squares which remove the singular fade subspaces for the case when the nodes have equal number of antennas $n=\max \lbrace n_A, n_B \rbrace$(Section III B).
\item
The singular fade subspaces which are removed by the conventional Exclusive-OR map are identified (Section III C).
\item
It is shown that finding the network coding maps which remove all the singular fade subspaces reduces to finding a small set of maps. The entire set can be obtained from the small set by the notions of isotopic and transposed Latin Squares (Section III C).
\item
The set of all Latin Squares which remove all the singular fade subspaces for the case when $n_A=n_B=2$ and QPSK signal set is used at the end nodes is explicitly provided (Section IV).
\item
It is shown that most of the Latin Squares which remove the singular fade subspaces for the $n \times n$ system, $n \geq 2,$ are obtainable from Latin Squares which remove the singular fade subspaces of the $m \times m$ system, where $m<n$ (Section V).  
\end{itemize}

\section{SINGULAR FADE SUBSPACES FOR $2^\lambda$-PSK SIGNAL SET}
In this section, the structure as well as the total number of singular fade subspaces is obtained for arbitrary $2^\lambda$-PSK signal sets.  The points in the symmetric $M$-PSK signal set are assumed to be of the form $e^{j (2k+1) \pi/M},0 \leq k \leq M-1$ and $M$ is of the form $2^\lambda$, where $\lambda$ is a positive integer.
 
For any $M$-PSK signal set, the set $\Delta\mathcal{S}$ is of the form,

{\footnotesize
\begin{align}
\nonumber
\Delta\mathcal{S}=&\left\lbrace 0\right\rbrace\cup \left\lbrace 2\sin(\pi n /M) e^{j k 2 \pi/M}  \vert{n \; \textrm{odd} }\right\rbrace\\
\label{psk_diff}
&\hspace{2 cm}\cup\left\lbrace 2\sin(\pi n /M) e^{j (k 2 \pi/M +  \pi/M)}\vert{n \; \textrm{even} }\right\rbrace,
\end{align}
}where $1 \leq n \leq M/2$ and $0 \leq k \leq M-1$.

\begin{figure}[htbp]
\centering
\includegraphics[totalheight=3.3in,width=6in]{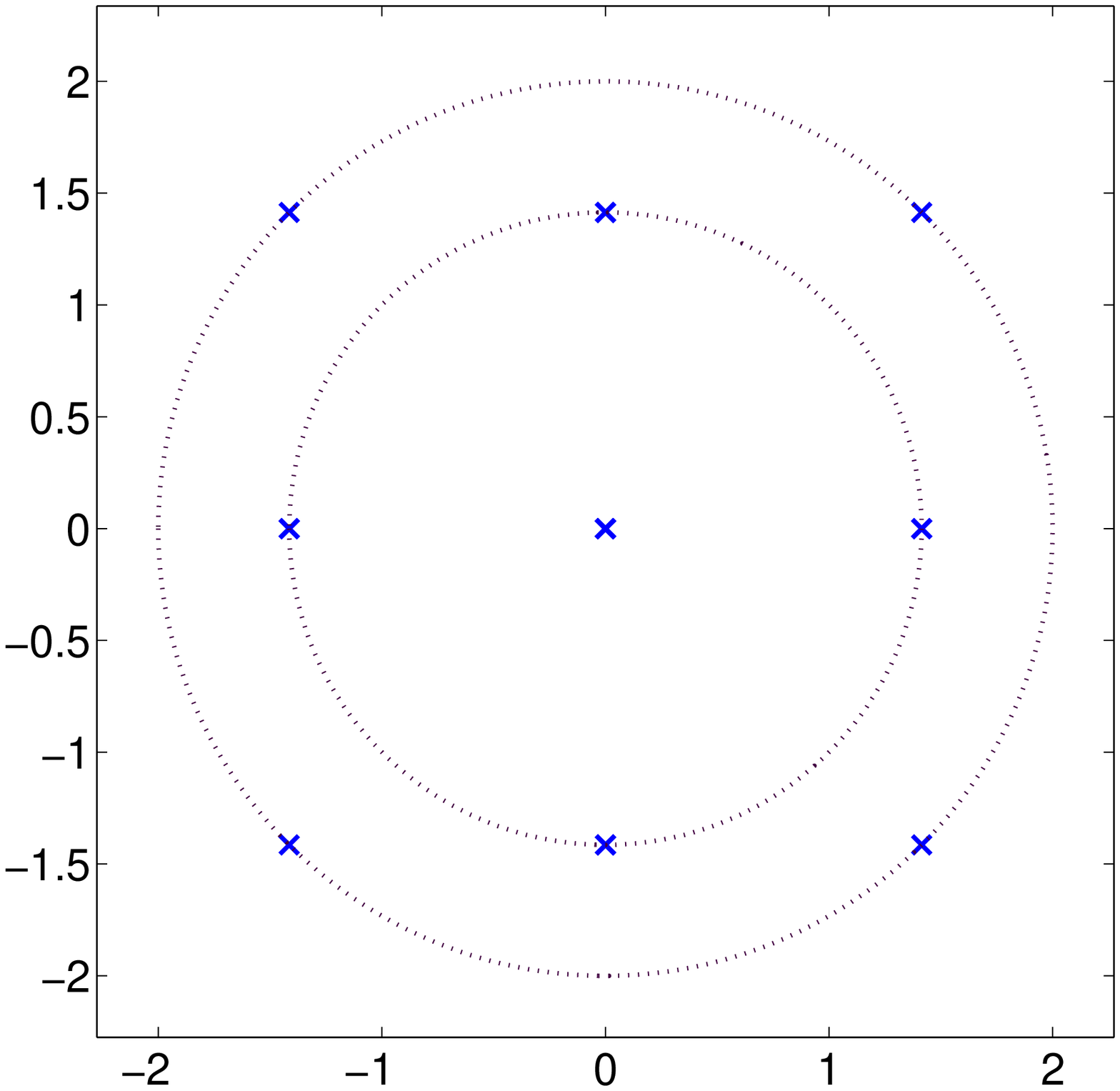}
\caption{Difference constellation for QPSK signal set}	
\label{4psk_diff}	
\end{figure}

In other words, the non-zero points in $\Delta\mathcal{S}$ lie on $M/2$ circles of radius $2\sin(\pi n/M), 1 \leq n \leq M/2$ with each circle containing $M$ points. The phase angles of the $M$ points on each circle is $2 k \pi/M$, if $n$ is odd and $2k \pi/M+\pi/M$ if $n$ is even, where $0 \leq k \leq M-1$. For example the difference constellation for QPSK signal set is shown in Fig. \ref{4psk_diff}.

%
%\begin{figure}[htbp]
%\centering
%\includegraphics[totalheight=2.75in,width=5in]{8psk_diff}
%\caption{Difference constellation for 8-PSK signal set}	
%\label{8psk_diff}	
%\end{figure}

%Let us define,
%
%\begin{align}
%\nonumber
%x_{k,n}=\left\lbrace
%\begin{array}{ll}
%\nonumber
%2\sin(\pi n /M) e^{j k 2 \pi/M} {\;\textrm{if}\; n \; \textrm{is} \; \textrm{odd}, }\\
%2\sin(\pi n /M) e^{j (k 2 \pi/M +  \pi/M)} {\;\textrm{if}\; n \; \textrm{is} \; \textrm{even}, }
%\end{array}
%\right.\\
%\label{eqn_diff}
%\end{align}
%where $1 \leq n \leq M/2$ and $0 \leq k \leq M-1$.
%
%From \eqref{eqn_dmin}, it follows that the singular fade states are of the form, 
%
%{\vspace{-.6 cm}
%\begin{align*}
%\gamma_{s} e^{j \theta_{s}}=-x_{k,n}/x_{k',n'}, 
%\end{align*}
%}for some $x_{k,n},x_{k',n'} \in \Delta\mathcal{S}$.

The following Lemma is useful in finding the structure as well as number of singular fade subspaces. 

\begin{lemma}
\label{lemma_trig}
For integers $k_1$, $k_2$, $l_1$ and $l_2$, where $$1 \leq k_1,k_2,l_1,l_2 \leq \frac{M}{2},k_1 \neq k_2  \textrm{ and }  l_1 \neq l_2,$$

{\footnotesize
\vspace{-0.4 cm}
\begin{align}
\label{eqn_lemma_trig}
\dfrac{\sin(k_1 \pi/M)}{\sin(k_2 \pi/M)}=\dfrac{\sin(l_1 \pi/M)}{\sin(l_2 \pi/M)},
\end{align}
}if and only if $k_1 = l_1$ and  $k_2 = l_2$.
\begin{proof}
See \cite{VNR}.
\end{proof}
\end{lemma}

Recall from Section I that the singular fade subspaces are of the form  $[\mathrm{span}( [\Delta x_A^T \; \Delta x_B^T]^T)]^\perp,$ where $\Delta x_A \in \Delta \mathcal{S}^{n_A}$ and $\Delta x_B \in \Delta \mathcal{S}^{n_B}.$  Let $\Delta x_A ^{m}$ ($\Delta x_B ^{m}$) denote the $m^{th}$ element of $\Delta x_A$ ($\Delta x_B$). Let $ i_1,i_2,...,i_L $ be the ordered indices for which $\Delta x_{A}^{i_k} \neq 0, 1 \leq k \leq L.$ Similarly, let $ j_1,j_2,...,j_{L'}$ be the ordered indices for which $\Delta x_{B}^{j_k} \neq 0, 1 \leq k \leq L'.$ Let $\phi^{i_k}_{A}, 1 \leq k \leq L$ and  $\phi^{j_l}_{B}, 1 \leq l \leq L'$ denote the phase angles of $\Delta x^{i_k}_{A}$ and $\Delta x^{j_l}_{B}$ respectively. Let the vector of length $L+L'-1,$ 
\begin{align*}
&[(\phi_{A}^{i_2}-\phi_{A}^{i_1})\quad (\phi_{A}^{i_3}-\phi_{A}^{i_1}) ... (\phi_{A}^{i_L}-\phi_{A}^{i_1})\\
&\hspace{2 cm}(\phi_{B}^{j_1}-\phi_{A}^{i_1})\quad(\phi_{B}^{j_2}-\phi_{A}^{i_1}) ... (\phi_{B}^{j_{L'}}-\phi_{A}^{i_1})]^T
\end{align*}
 be referred to as the relative phase vector of $\Delta x.$

Note that $[\mathrm{span}(\Delta x)]^\perp$ and $[\mathrm{span}(\Delta x')]^\perp$ can be the same for some $\Delta x, \Delta x' \in \Delta \mathcal{S}^{n_A+n_B}.$ A necessary and sufficient condition for $[\mathrm{span}(\Delta x)]^\perp=[\mathrm{span}(\Delta x')]^\perp,$ is that $\mathrm{span}(\Delta x)=\mathrm{span}(\Delta x'),$ i.e., $\Delta x= c \Delta x',$ for some $c \in \mathbb{C}.$ Equivalently, the conditions for $[\mathrm{span}(\Delta x)]^\perp=[\mathrm{span}(\Delta x')]^\perp$can be stated as follows:
\begin{itemize}
\item
The location of the non-zero components is the same in the vectors $\Delta x$ and $\Delta x'.$
\item
 The relative phase vector of $\Delta x$ and $\Delta x'$ are equal. 
\item
$\vert \Delta x_i \vert=c\vert \Delta x'_i \vert,$ $\forall 1 \leq i \leq n_A+n_B,$ for some $c \in \mathbb{C}.$  
\end{itemize}
%The following lemma identifies the exact conditions under which it happens for $2^\lambda$-PSK signal set. 
For $2^\lambda$ PSK signal set, given the first condition, the third condition given above can be replaced by the condition given in the following lemma. 
\begin{lemma}
\label{lemma_same_space}
Let $i_1,i_2,i_3,...,i_L$ be the ordered indices corresponding to the non-zero components in $\Delta x$ and $\Delta x'$ (the location of the non-zero components is the same in the vectors $\Delta x$ and $\Delta x'$). For $2^{\lambda}$-PSK signal set, $\vert \Delta x_i \vert=c\vert \Delta x'_i \vert,$ $\forall 1 \leq i \leq n_A+n_B,$ for some $c \in \mathbb{C},$ if and only if the magnitude of the non-zero components in $\Delta x$ are equal and the magnitudes of the non-zero components in $\Delta x'$ are equal, i.e., $\vert \Delta x_{i_1} \vert = \vert \Delta x_{i_2} \vert=...\vert \Delta x_{i_L}\vert$ and $\vert \Delta x'_{i_1} \vert = \vert \Delta x'_{i_2} \vert=...\vert \Delta x'_{i_L}\vert.$
\begin{proof}
When the condition given in the statement of the lemma is satisfied, clearly, $\vert \Delta x_i \vert=c\vert \Delta x'_i \vert,$ $\forall 1 \leq i \leq n_A+n_B.$

The proof of the ``only if" part is as follows. Let $1 \leq r_1<r_2<...<r_L \leq n_A$ be the indices for which $\Delta x_{A}^{r_k} \neq 0, \Delta {x'}_{A}^{r_k} \neq 0, 1 \leq k \leq L.$ Similarly, let $1 \leq j_1<j_2<...<j_L' \leq n_B$ be the indices for which $\Delta x_{B}^{j_k} \neq 0, \Delta {x'}_{B}^{j_k} \neq 0, 1 \leq k \leq L'.$ Since $\mathrm{span}(\Delta x)=\mathrm{span}(\Delta x'),$ we have,

{\scriptsize
\begin{equation}
\label{eqn_modulus}
\frac{\Delta x_{A}^{r_1}} {\Delta {x'}_{A}^{r_1}}=\frac{\Delta x_{A}^{r_2}} {\Delta {x'}_{A}^{r_2}}=...=\frac{\Delta x_{A}^{r_L}} {\Delta {x'}_{A}^{r_L}}=\frac{\Delta x_{B}^{j_1}}{\Delta {x'}_{B}^{j_1}}=\frac{\Delta x_{B}^{j_2}} {\Delta {x'}_{B}^{j_2}}=...=\frac{\Delta x_{B}^{j_L'}} {\Delta {x'}_{B}^{j_L'}}.
\end{equation}
}

From \eqref{psk_diff} it follows that the absolute value of the ratio of the points in the difference constellation $\Delta \mathcal{S}$ are of the form $\frac{\sin(\frac{k\pi}{M})}{\sin(\frac{l\pi}{M})}, \mathrm{ for}\; \mathrm{some} \: 1 \leq k,l \leq M/2.$ Equating the absolute values of the terms in \eqref{eqn_modulus} and from Lemma \ref{lemma_trig}, it follows that the absolute values of the non-zero components in $\Delta x$ need to be equal and the absolute values of the non-zero components in $\Delta x'$ need to be equal. This completes the proof.
%Equating the phase angles of the terms in \eqref{eqn_modulus}, we have 
%
%{\small
%\begin{align*}
%\phi_{A_{i_1}}-\phi'_{A_{i_1}}&=\phi_{A_{i_2}}-\phi'_{A_{i_2}}=...=\phi_{A_{i_L}}-\phi'_{A_{i_L}}\\
%&\hspace{-.2cm}=\phi_{B_{j_1}}-\phi'_{B_{j_1}}=\phi_{B_{j_2}}-\phi'_{B_{j_2}}=...=\phi_{B_{j_L'}}-\phi'_{B_{j_L'}}.
%\end{align*}
%}
%Equivalently, the above condition can be written as
%
%
%{\small
%\begin{align*}
%&\phi_{A_{i_2}}-\phi_{A_{i_1}}=\phi_{A'_{i_2}}-\phi_{A'_{i_1}},...,\phi_{A_{i_L}}-\phi_{A_{i_L}}=\phi_{A'_{i_L}}-\phi_{A'_{i_L}},\\
%&\phi_{B_{j_1}}-\phi_{A_{i_1}}=\phi_{B'_{j_1}}-\phi_{A'_{i_1}},...,\phi_{A_{i_L}}-\phi_{A_{i_1}}=\phi_{B'_{j_L'}}-\phi_{A'_{i_1}}.
%\end{align*}
%}
%\noindent This completes the proof.

\end{proof}
\end{lemma}

\begin{example}
For the $2 \times 2$ system with BPSK signal set, for $\Delta x =[0 \:\:2 \:\: 2 \: -2]^T$ and $\Delta x' =[0 \:-2 \:-2 \:\: 2]^T,$ the location of the non-zero components in $\Delta x$ and $\Delta x'$ are the same. The relative phase vectors of $\Delta x$ and $\Delta x'$ are equal. Also, we have, $\vert \Delta x_2 \vert=\vert \Delta x_3 \vert=\vert \Delta x_4 \vert=2,\vert \Delta x'_2 \vert=\vert \Delta x'_3 \vert=\vert \Delta x'_4 \vert=2$. Hence, $[\text{span}([0 \:\:2 \:\: 2 \: -2]^T)]^\perp=[\text{span}([0 \:-2 \:- 2 \:\: 2]^T)]^\perp.$
\end{example}

\begin{example}
For the $2 \times 2$ system with QPSK signal set, it can be seen from Fig. \ref{4psk_diff} that the non-zero points in the difference constellation lie on circles with radii $\sqrt{2}$ and 2. For $\Delta x =[0 \quad \sqrt{2}+j\sqrt{2} \quad \sqrt{2}-j\sqrt{2} \quad \sqrt{2}+j\sqrt{2}]^T)]$ and $\Delta x' =[0 \quad \sqrt{2} \quad-j\sqrt{2} \quad \sqrt{2}]^T)],$  the location of the non-zero components in $\Delta x$ and $\Delta x'$ are the same. The relative phase vectors of $\Delta x$ and $\Delta x'$ are equal. Also, we have, $\vert \Delta x_2 \vert=\vert \Delta x_3 \vert=\vert \Delta x_4 \vert=4,\vert \Delta x'_2 \vert=\vert \Delta x'_3 \vert=\vert \Delta x'_4 \vert=2.$ Hence the vector subspaces  $[\text{span}([0 \quad \sqrt{2}+j\sqrt{2} \quad \sqrt{2}-j\sqrt{2} \quad \sqrt{2}+j\sqrt{2}]^T)]^\perp$ and $[\text{span}([0 \quad \sqrt{2} \quad-j\sqrt{2} \quad \sqrt{2}]^T)]^\perp$ are the same.
\end{example}
%\begin{example}
%Consider the case when $4$-PSK signal set is used at the end nodes, with $n_A=n_R=n_B=2.$ The difference constellation $\Delta \mathcal{S}$ has 8 non-zero points which lie on circles with radii $\sqrt{2}$ and $2.$ The 4 points on the circle with radius $\sqrt{2}$ have phase angles $\frac{2k\pi}{4}, k \in \lbrace 0,1,2,3 \rbrace$ and the 4 points on the circle with radius ${2}$ have phase angles $\frac{(2k+1)\pi}{4}, k \in \lbrace 0,1,2,3 \rbrace$.
%\end{example}

As mentioned in Section I, the singular fade subspaces can be classified in to removable and non-removable singular fade subspaces. The non-removable singular fade subspaces are of the form $[\mathrm{span}([\Delta x_A \; 0_{n_B}]^T)]^\perp, \Delta x_A \in \Delta \mathcal{S}^{n_A}$ or $[\mathrm{span}([ 0_{n_A} \; \Delta x_B]^T)]^\perp, \Delta x_B \in \Delta \mathcal{S}^{n_B}.$ The removable singular fade subspaces are of the form $\left[\mathrm{span}([\Delta x_A^T \; \Delta x_B^T]^T)\right]^\perp, \Delta x_A \neq 0_{n_A}, \Delta x_B \neq 0_{n_B}.$ 

The following lemma gives the total number of non-removable and removable singular fade subspaces.

\begin{lemma}
\label{sing_non_rem}
For $M$-PSK signal set ($M$ any power of 2), 

\begin{itemize}
\item
the total number of non-removable singular fade subspaces is given by,

{\footnotesize 
\begin{align*}
&\sum_{k=1}^{n_A} {n_A \choose k}  \left[\left(\frac{M}{2}\right)^{k}-\frac{M}{2}+1 \right]M^{k-1}\\
&\hspace{2 cm}+\sum_{l=1}^{n_B} {n_B \choose l}  \left[\left(\frac{M}{2}\right)^{l}-\frac{M}{2}+1 \right]M^{l-1}.
\end{align*}
}
\item
 the number of removable singular fade subspaces of the form $[\text{span}(\Delta x)]^\perp$ with $2 \leq k \leq {n_A+n_B}$ non-zero components in $\Delta x$ is given by,

{\footnotesize $$\left[{n_A+n_B \choose k}-{n_A \choose k}-{n_B \choose k}\right] \left[\left(\frac{M}{2} \right)^{k}-\frac{M}{2}+1 \right]M^{k-1}.$$} Hence the total number of removable singular fade subspaces is,

{\footnotesize
\begin{align*}
&\sum_{k=2}^{n_A+n_B} \left(\left[{n_A+n_B \choose k}-{n_A \choose k}-{n_B \choose k}\right] \right.\\
&\hspace{4 cm}\left.\left[\left(\frac{M}{2} \right)^{k}-\frac{M}{2}+1 \right]M^{k-1}\right),
\end{align*}
}where $a \choose b$ is defined to be zero if $b>a.$
\end{itemize}
\begin{proof}
See Appendix A.
\end{proof}
\end{lemma}
\begin{example}
Consider the MIMO two-way relaying system with $n_A=2$ and $n_B=2.$ From Lemma \ref{sing_non_rem}, the non-removable singular fade subspaces for BPSK are 4 in number and are the ones numbered from 1 to 4 in Table I. For QPSK signal set, there are 28 non-removable singular fade subspaces. For BPSK and QPSK signal sets, the number of removable singular fade subspaces of the form $[\text{span}(\Delta x)] ^\perp$, where $\Delta x$ has $k$ non-zero entries are as given below: 

\vspace{-0.5 cm}
{\centering
$$\begin{array}{|c|c|c|}
\cline{1-3} \multicolumn{1}{|c|}{k} & \multicolumn{2}{c|}{\text{No. of removable singular fade subspaces}}\\ \cline{2-3}
 \multicolumn{1}{|c|}{}&\multicolumn{1}{c|}{\textsc{~~~~~~~~~~BPSK~~~~~~~~~~}}& \multicolumn{1}{c|}{\textsc{QPSK}}\\
\cline{1-3} 2   & 8 &48\\
\cline{1-3} 3   & 16 &448\\
\cline{1-3} 4   & 8 &960\\
\cline{1-3} \text{Total} & 32 & 1456\\
\cline{1-3}
\end{array}$$
}
The removable singular fade subspaces for BPSK signal set, which are 32 in number, are the ones numbered from 5 to 36 in Table I. 
\end{example}

{
\begin{table}
\centering
\footnotesize
\begin{tabular}{|c|r@{}l|r@{}l}
\hline \textbf{No.} & \textbf{Singular fade subspace}& \\
%\hline 1   & $[\text{span}([0 \:0 \: 0\: 0]^T)]^\perp$&   \\
\hline 1   & $[\text{span}([0 \:0 \: 2\: 2]^T)]^\perp$&=$[\text{span}([0 \:0 \: {-2}\: {-2}]^T)]^\perp$  \\
\hline 2   & $[\text{span}([0 \:0 \: 2\: {-2}]^T)]^\perp$&=$[\text{span}([0 \:0 \: {-2}\: 2]^T)]^\perp$  \\
\hline 3   & $[\text{span}([2 \:2 \: 0\: {0}]^T)]^\perp$&=$[\text{span}([{-2} \:{-2} \: {0}\: 0]^T)]^\perp$ \\
\hline 4   & $[\text{span}([{2} \:{-2} \: 0\: {0}]^T)]^\perp$&=$[\text{span}([{-2} \:{2} \: {0}\: 0]^T)]^\perp$    \\
\hline 5 & $[\text{span}([0 \:{2} \: 0\: {2}]^T)]^\perp$&=$[\text{span}([{0} \:{-2} \: {0}\: {-2}]^T)]^\perp$  \\
\hline 6 & $[\text{span}([0 \:{2} \: 0\: {-2}]^T)]^\perp$&=$[\text{span}([{0} \:{-2} \: {0}\: {2}]^T)]^\perp$  \\
\hline 7 & $[\text{span}([2 \:{0} \: {2}\: {0}]^T)]^\perp$&=$[\text{span}([{-2} \:{0} \: {-2}\: {0}]^T)]^\perp$ \\
\hline 8 & $[\text{span}([2 \:{0} \: {-2}\: {0}]^T)]^\perp$&=$[\text{span}([{-2} \:{0} \: {2}\: {0}]^T)]^\perp$ \\
\hline 9 & $[\text{span}([ {2} \: 0 \: 0\: {2}]^T)]^\perp$&=$[\text{span}([{-2} \: 0  \: {0}\: {-2}]^T)]^\perp$ \\
\hline 10 & $[\text{span}([2 \:{0} \: 0\: {-2}]^T)]^\perp$&=$[\text{span}([{-2} \: 0 \: {0}\: {2}]^T)]^\perp$  \\
\hline 11 & $[\text{span}([0 \:{2} \: 2 \: {0}]^T)]^\perp$&=$[\text{span}([{0} \:{-2} \: {-2} \:{0}]^T)]^\perp$ \\
\hline 12 & $[\text{span}([0 \:{2} \: {-2} \:{0}]^T)]^\perp$&=$[\text{span}([{0} \:{-2} \: 2 \:{0}]^T)]^\perp$    \\
\hline 13   & $[\text{span}([0 \:2 \: 2\: {2}]^T)]^\perp$&=$[\text{span}([0 \:{-2} \: {-2}\: {-2}]^T)]^\perp$    \\
\hline 14   & $[\text{span}([0 \:2 \: 2\: {-2}]^T)]^\perp$&=$[\text{span}([0 \:{-2} \: {-2}\: {2}]^T)]^\perp$  \\
\hline 15   & $[\text{span}([0 \:2 \: {-2}\: {-2}]^T)]^\perp$&=$[\text{span}([0 \:{-2} \: {2}\: {2}]^T)]^\perp$  \\
\hline 16   & $[\text{span}([0 \:2 \: {-2}\: {2}]^T)]^\perp$&=$[\text{span}([0 \:{-2} \: {2}\: {-2}]^T)]^\perp$  \\
\hline 17   & $[\text{span}([2 \: 0 \: 2\: {2}]^T)]^\perp$&=$[\text{span}([{-2} \: 0 \: {-2}\: {-2}]^T)]^\perp$  \\
\hline 18   & $[\text{span}([2 \: 0 \: 2\: {-2}]^T)]^\perp$&=$[\text{span}([{-2}\: 0  \: {-2}\: {2}]^T)]^\perp$ \\
\hline 19   & $[\text{span}([2 \: 0 \: {-2}\: {-2}]^T)]^\perp$&=$[\text{span}([{-2} \: 0 \: {2}\: {2}]^T)]^\perp$\\
\hline 20   & $[\text{span}([2 \: 0 \: {-2}\: {2}]^T)]^\perp$&=$[\text{span}([{-2} \: 0 \: {2}\: {-2}]^T)]^\perp$ \\
\hline 21   & $[\text{span}([2 \: 2\: 0 \: {2}]^T)]^\perp$&=$[\text{span}([{-2} \: {-2} \:{0}\: {-2}]^T)]^\perp$ \\
\hline 22 &$[\text{span}(2 \: 2\: 0 \: {-2}]^T)]^\perp$&=$[\text{span}([{-2} \: {-2}\:0\: {2}]^T)]^\perp$\\
\hline  23 & $[\text{span}([2 \: {-2}\:0\: {-2}]^T)]^\perp$&=$[\text{span}([{-2} \: {2}\:0\: {2}]^T)]^\perp$\\
\hline 24 &  $[\text{span}([2 \: {-2}\:0\: {2}]^T)]^\perp$&=$[\text{span}([{-2} \: {2}\:0\: {-2}]^T)]^\perp$\\
\hline   25   & $[\text{span}([2 \: 2 \: {2}\: 0]^T)]^\perp$&=$[\text{span}([{-2} \: {-2} \: {-2}\: 0]^T)]^\perp$\\
\hline 26 &$[\text{span}(2 \: 2 \: {-2} \: 0]^T)]^\perp$&=$[\text{span}([{-2} \: {-2}\: {2} \: 0]^T)]^\perp$\\
\hline  27 & $[\text{span}([2 \: {-2}\: {-2} \:0]^T)]^\perp$&=$[\text{span}([{-2} \: {2}\: {2} \:0]^T)]^\perp$ \\
\hline 28 &  $[\text{span}([2 \: {-2}\: {2} \:0]^T)]^\perp$&=$[\text{span}([{-2} \: {2}\: {-2} \:0]^T)]^\perp$ \\
\hline 29 &  $[\text{span}([{2} \: {2}\: {2} \:{2}]^T)]^\perp$&=$[\text{span}([{-2} \: {-2}\: {-2} \:{-2}]^T)]^\perp$\\
\hline 30 &  $[\text{span}([{2} \: {2}\: {2} \:{-2}]^T)]^\perp$&=$[\text{span}([{-2} \: {-2}\: {-2} \:{2}]^T)]^\perp$\\
\hline 31 & $[\text{span}([{2} \: {2}\: {-2} \:{-2}]^T)]^\perp$&=$[\text{span}([{-2} \: {-2}\: {2} \:{2}]^T)]^\perp$\\
\hline 32 &  $[\text{span}([{2} \: {2}\: {-2} \:{2}]^T)]^\perp$&=$[\text{span}([{-2} \: {-2}\: {2} \:{-2}]^T)]^\perp$\\
\hline 33 &  $[\text{span}([{2} \: {-2}\: {2} \:{2}]^T)]^\perp$&=$[\text{span}([{-2} \: {2}\: {-2} \:{-2}]^T)]^\perp$ \\
\hline  34 &  $[\text{span}([{2} \: {-2}\: {2} \:{-2}]^T)]^\perp$&=$[\text{span}([{-2} \: {2}\: {-2} \:{2}]^T)]^\perp$  \\
\hline  35 &  $[\text{span}([{2} \: {-2}\: {-2} \:{2}]^T)]^\perp$&=$[\text{span}([{-2} \: {2}\: {2} \:{-2}]^T)]^\perp$ \\
\hline 36 &  $[\text{span}([{2} \: {-2}\: {-2} \:{-2}]^T)]^\perp$&=$[\text{span}([{-2} \: {2}\: {2} \:{2}]^T)]^\perp$ \\
\hline
\end{tabular}
\label{table1}
\caption{singular fade subspaces for the $2 \times 2$ system with BPSK signal set, where the singular fade subspaces numbered 1 to 4 are non-removable and the rest are removable}
\end{table}
}

From Lemma \ref{sing_non_rem}, it can be seen that the number non-removable singular fade subspaces is $O(M^{2\max\lbrace n_A,n_B\rbrace -1}),$  while the number of removable singular fade subspaces is $O(M^{2(n_A+n_B)-1}).$ Hence, the number of non-removable singular fade subspaces is a small fraction of the total number of  singular fade subspaces and the fraction tends to zero for increasing values of $M.$
\section{The Exclusive Law and Latin Rectangles}
For the two-way relaying scenario, with single antenna at the nodes, with signal sets of equal cardinality used at the end nodes, it was shown in \cite{NVR} that all network coding maps satisfying the exclusive law are representable as Latin Squares. In this section, we establish the connection between Latin Rectangles and network coding maps satisfying the exclusive law, for the MIMO two-way relaying scenario.
  
\textit{Definition 4:} \cite{Sto} A Latin Rectangle L of order $N_1 \times N_2$ on the symbols from the set $\mathbb{Z}_t=\{0,1, \cdots ,t-1\}$ is an ${N_1} \times {N_2}$  array, in which each cell contains one symbol and each symbol occurs at most once in each row and column. A Latin Rectangle of order $N \times N$ is called a Latin Square of order $N.$ 

 Let the points in the $M$-point signal set used for transmission at the nodes be indexed by the elements of the set $\mathbb{Z}_{M}={\lbrace 0,1,2,\dotso,M-1 \rbrace}.$
Consider an $ M^{n_A} \times M^{n_B}$ array at the relay with the rows ($\setminus$columns) indexed by the $n_A$-tuple($\setminus n_B$-tuple) $[x_{A_1},x_{A_2},\dotso,x_{A_{n_A}}]$ ($\setminus [x_{B_1},x_{B_2},\dotso,x_{B_{n_B}}]$) denoting the complex vector transmitted by node A ($\setminus$B). Our aim is to form clusters from the slots in the $M^{n_A} \times M^{n_B}$ array such that the exclusive law is satisfied. To do so, we will fill in the slots in the array with the elements of set $\mathbb{Z}_{t}$  and the clusters are obtained by taking all the row-column pairs $(i,j),i \in \mathbb{Z}_{M}^{n_A}, j \in \mathbb{Z}_{M}^{n_B}$ such that the entry in the $(i,j)-$th slot is the same symbol from  $\mathbb{Z}_{t}$ for a cluster. The specific symbols from $\mathbb{Z}_t$ are not important, but it is the set of clusters that are important. 
Now, it is easy to see that if the exclusive law need to be satisfied, then the clustering should be such that an element in a row (column) cannot be repeated in the same row (column). Thus all the relay clusterings which satisfy the exclusive law form Latin Rectangles. Hence, we have the following:

\textit{For MIMO two-way relaying, every relay clustering which satisfy the exclusive law forms a Latin Rectangle and vice verse.}

With this observation, the study  of clustering which satisfies the exclusive law can be equivalently carried out as the study  of  Latin Rectangles with appropriate parameters.

%%%%%%%%%%%%%%%% 
\subsection{Removing singular fade subspaces, Singularity-removal Constraints and Constrained Latin Rectangles}

%The relay can manage with constellations of size $M$ in BC phase, but it is observed that in some cases relay may not be able to remove the singular fade states with $t=M$ and results in severe performance degradation in the MA phase \cite{APT1}. 
Consider a singular fade subspace $f \in \mathcal{F}.$ Let $(k,l)(k^{\prime},l^{\prime}) \in \mathbb{Z}_{M}^{n_A} \times \mathbb{Z}_{M}^{n_B}$ be the pairs which result in $\Delta x$ such that $[\mathrm{span}(\Delta x)]^\perp=f$. If $(k,l)$ and $(k^{\prime},l^{\prime})$ are not clustered together, the minimum cluster distance will be zero, for all $H_R$ such that $\mathcal{R}(H_R) \preceq f.$ To avoid this, those pairs should be in the same cluster. This requirement is termed as {\it singularity-removal constraint}. So, we need to obtain Latin Rectangles which can remove singular fade subspaces and with minimum value for $t.$ Therefore, initially we will fill the slots in the  $M^{n_A} \times  M^{n_B}$ array  such that for the slots corresponding to a singularity-removal constraint the same element will be used to fill slots. This removes that particular singular fade subspace. Such a partially filled Latin Rectangle is called a {\it Constrained Partial Latin Rectangle} (CPLR). After this,  to make this a Latin Rectangle, we will try to fill the other slots of the  partially filled CPLR with minimum number of symbols from the set $\mathbb{Z}_{t}.$

\begin{example}
Consider the $2 \times 2$ system with BPSK signal set used at the end nodes. Consider the singular fade subspace $[\mathrm{span}([2 \:\:\: 2 \:\: -2 \: -2]^T)]^\perp=[\mathrm{span}([-2 \: -2 \:\:\: 2 \:\:\:2]^T)]^\perp.$  The singularity-removal constraint for this singular fade subspace is $\lbrace ([0 \; 0],[ 1 \;1]),([1 \; 1],[0 \; 0])\rbrace.$ The constrained partial Latin Square for this case is shown in Fig. \ref{CPLR_1}. The clustering which removes this singular fade subspace, given in \eqref{clustering1} in Example 3, can also be represented as a Latin Square shown in Fig. \ref{lat_1}. For the singular fade subspace $[\mathrm{span}([0 \: -2 \: -2 \:\:\:2]^T)]^\perp,$ the singularity removal constraints are $\left \lbrace \left \lbrace \left(\left[0 
\:0\right],\left[0 \: 1\right]\right),\left(\left[0 \:1\right],\left[1 \: 0\right]\right) \right \rbrace , \left \lbrace  \left(\left[1 
\:0\right],\left[0 \: 1\right]\right),\left(\left[1 \:1\right],\left[1 \: 0\right]\right) \right \rbrace \right \rbrace.$ The constrained partial Latin Square and the filled Latin Square which removes this singular fade subspace are shown respectively in Fig. \ref{CPLR_2} and Fig. \ref{lat_2}.

\begin{figure}
\centering
\begin{tabular}{|c|c|c|c|c|}
\hline  & 00 & 01 & 10 & 11\\
\hline 00 &  &  &  & 3\\
\hline 01 &  & &  & \\
\hline 10 &  &  &  & \\
\hline 11 & 3 &  &  & \\
\hline
\end{tabular}
\caption{Constrained Partial Latin Square corresponding to the singular fade subspace $[\mathrm{span}([2 \:\:\: 2 \:\: -2 \: -2]^T)]^\perp$}
\label{CPLR_1}
\end{figure}

\begin{figure}
\centering
\begin{tabular}{|c|c|c|c|c|}
\hline  & 00 & 01 & 10 & 11\\
\hline 00 & 0 & 1 & 2 & 3\\
\hline 01 & 1 & 0 & 3 & 2\\
\hline 10 & 2 & 3 & 0 & 1\\
\hline 11 & 3 & 2 & 1 & 0 \\
\hline
\end{tabular}
\caption{Latin Square corresponding to the bit-wise XOR map for BPSK signal set}
\label{lat_1}
\end{figure}

\begin{figure}
\centering
\begin{tabular}{|c|c|c|c|c|}
\hline  & 00 & 01 & 10 & 11\\
\hline 00 &  & 0 &  & \\
\hline 01 &  &  & 0 & \\
\hline 10 &  & 1 &  & \\
\hline 11 &  &  & 1 & \\
\hline
\end{tabular}
\caption{Constrained Partial Latin Square corresponding to the singular fade subspace $[\mathrm{span}([0 \: -2 \: -2 \:\:\:2]^T)]^\perp.$}
\label{CPLR_2}
\end{figure}

\begin{figure}
\centering
\begin{tabular}{|c|c|c|c|c|}
\hline  & 00 & 01 & 10 & 11\\
\hline 00 & 3 & 0 & 2 & 1\\
\hline 01 & 1 & 2 & 0 & 3\\
\hline 10 & 0 & 1 & 3 & 2\\
\hline 11 & 2 & 3 & 1 & 0 \\
\hline
\end{tabular}
\caption{Latin Square that removes the singular fade subspace $[\mathrm{span}([0 \: -2 \: -2 \:\:\:2]^T)]^\perp.$}
\label{lat_2}
\end{figure}
\end{example}

%%%%%%%%%%%%%%%%%%%%%%%%%%%%%%%%%%%%%%%%%%%%%%%%%%%%%
\subsection{Obtaining Latin Rectangles from Latin Squares}
In this section, it is shown that the Latin Rectangles which remove the singular fade subspaces for the $n_A  \times n_B$ system can be obtained from the Latin Squares,  which remove the singular fade subspaces for the $n  \times n$ system, where $n=\max \lbrace n_A,n_B \rbrace,$ by removing certain rows or columns. 

For $n_B > n_A,$ if $[\text{span}([\Delta x_A^T \; \Delta x_B^T]^T)]^\perp$ is a singular fade subspace for the $n_A  \times n_B$ system, then $[\text{span}([0_{n_B-n_A}^T \; \Delta {x_A^T} \; \Delta x_B^T]^T)]^{\perp}$ is a singular fade subspace for the $n_B  \times n_B$ system. Similarly, for $n_A >n_B$, if $[\text{span}([\Delta x_A^T \; \Delta x_B^T]^T)]^{\perp}$ is a singular fade subspace for the $n_A  \times n_B$ system, then $[\text{span}([\Delta {x_A^T} \; 0_{n_A-n_B}^T \; \Delta {x_B^T}]^T)]^{\perp}$ is a singular fade subspace for the $n_A  \times n_A$ system.

For a Latin Square $L$ of order $n,$ let $L_{[1:r,:]}$ denote the Latin Rectangle of order $r \times n$ obtained by taking only the first $r$ rows of $L.$ Similarly, let $L_{[:,1:c]}$ denote the Latin Rectangle obtained by taking only the first $c$ columns of $L.$ 

%Assuming $n_B > n_A,$ the reason why the Latin Rectangle which removes the singular fade subspace for the $n_B \times n_R \times n_A$ system should be obtainable from a Latin Square which removes one of the singular fade subspaces for the $n_B \times n_R \times n_B$ system is as follows: The $n_A \times n_R \times n_B$ system can be viewed equivalently as the $n_B \times n_R \times n_B$ system, with $n_B-n_A$ antennas of $A$ (w.l.o.g. assumed to be the first $n_B-n_A$ antennas), always transmitting the same signal point (say the point in $\mathcal{S}$ index by 0) which is known to R. The above equivalence is valid since R can consider the transmission from the first $n_B-n_A$ antennas of A to be known interference and can cancel it perfectly. Since R knows perfectly the symbols transmitted by the first $n_B-n_A$ antennas, the only singular fade subspaces which can occur for the $n_B \times n_R \times n_B$ system are of the form $[\text{span}([0_{n_B-n_A}^T \; \Delta {x_A^T} \Delta x_B^T])]^{\perp},$ which can be viewed equivalently as the         
\begin{lemma}
\label{rect_from_square}
For $n_B > n_A,$ if the Latin Square $L$ removes the singular fade subspace $[\text{span}([0_{n_B-n_A}^T \; \Delta {x_A^T} \Delta x_B^T])]^{\perp}$ for the $n_B  \times n_B$ system, the Latin Rectangle  $L_{[1: M^{n_A},:]}$ removes the singular fade subspace $[\text{span}([\Delta {x_A^T} \; \Delta x_B^T])]^{\perp}$ for the $n_A \times n_B$ system. Similarly, for $n_A > n_B,$ if the Latin Square $L$ removes the singular fade subspace $[\text{span}([\Delta {x_A^T} \; 0_{n_A-n_B}^T \; \Delta {x_B^T}]^T)]^{\perp}$ for the $n_A  \times n_A$ system, the Latin Rectangle  $L_{[:,1:M^{n_B}]}$ removes the singular fade subspace $[\text{span}([\Delta {x_A^T} \; \Delta {x_B^T}]^T)]^{\perp}$ for the $n_A  \times n_B$ system. 
\begin{proof}
Consider the case when $n_B > n_A.$ Let $f$ denote the singular fade subspace $[\text{span}([0_{n_B-n_A}^T \; \Delta {x_A^T} \; \Delta x_B^T])]^{\perp}$ for the $n_B  \times n_B$ system. Let $\Delta \mathsf{x}_A= [ 0_{n_B-n_A}^T \Delta x_A ^T]^T.$ The singularity removal constraints for $f$ are of the form, 

\begin{align*}
\lbrace &([{\mathsf{x}_A^T}_{[1:n_B-n_A]} \; {\mathsf{x}_A^T}_{[n_B-n_A+1:n_B]}]^T, \mathsf{x}_B^T), \\
&\hspace{2 cm}([{\mathsf{x}_A^T}_{[1:n_B-n_A]} \; {\mathsf{x'}_B^T}_{[n_B-n_A+1:n_B]}, {\mathsf{x'}_B^T}) \rbrace,
\end{align*}
where ${\mathsf{x}_A}_{[n_B-n_A+1:n_B]}]-{\mathsf{x'}_B}_{[n_B-n_A+1:n_B]}=\Delta x_A,$  $\mathsf{x}_B-\mathsf{x'}_B=\Delta x_B$ and ${\mathsf{x}_A}_{[1:n_B-n_A]} \in  \mathcal{S} ^ {n_B-n_A}.$ For ${\mathsf{x}_A}_{[1:n_B-n_A]}= 0_{n_A-n_B},$ the singularity removal constraints are, 
\begin{align}
\nonumber
\lbrace &([0_{[1:n_B-n_A]} \; {\mathsf{x}_A^T}_{[n_B-n_A+1:n_B]}]^T, \mathsf{x}_B^T), \\
\label{zero_const}
&\hspace{2 cm}([0_{[1:n_B-n_A]} \; {\mathsf{x'}_B^T}_{[n_B-n_A+1:n_B]}, {\mathsf{x'}_B^T}) \rbrace.
\end{align}
All the cells given in the constraints \eqref{zero_const} belong to the rows 1 to $M n_A.$ Since the Latin Square $L$ which removes $[\text{span}([0_{n_B-n_A}^T \; \Delta {x_A^T} \; \Delta x_B^T])]^{\perp}$ satisfies the constraints in \eqref{zero_const}, the Latin Rectangle $L_{[1:M^{n_A},:]}$ satisfies the constraints 
\begin{align}
\nonumber
\lbrace &([{\mathsf{x}_A^T}_{[n_B-n_A+1:n_B]}]^T, \mathsf{x}_B^T),([{\mathsf{x'}_B^T}_{[n_B-n_A+1:n_B]}, {\mathsf{x'}_B^T}) \rbrace,
\end{align}
which are the singularity-removal constraints corresponding to the  singular fade subspace $[\text{span}([\Delta {x_A^T} \; \Delta x_B^T]^T)]^{\perp}$ for the $n_A  \times n_B$ system.
\end{proof}
\end{lemma}

\begin{example}
Consider the $1  \times 2$ system with BPSK signal set. The Latin Rectangle which removes the singular fade subspace $[\mathrm{span}([\: -2 \: -2 \:\:\:2]^T)]^\perp$ shown in Fig. \ref{latin_3} is obtained by taking only the first two rows of the Latin Square in Fig. \ref{lat_2} which removes the singular fade subspace $[\mathrm{span}([0 \: -2 \: -2 \:\:\:2]^T)]^\perp$ of the $2 \times 2$ system.
\begin{figure}
\centering
\begin{tabular}{|c|c|c|c|c|}
\hline  & 00 & 01 & 10 & 11\\
\hline 00 & 3 & 0 & 2 & 1\\
\hline 01 & 1 & 2 & 0 & 3\\
\hline
\end{tabular}
\caption{Latin Rectangle that removes the singular fade subspace $[\mathrm{span}([ -2 \: -2 \:\:\:2]^T)]^\perp.$}
\label{latin_3}
\end{figure}
\end{example}

From Lemma \ref{rect_from_square} it follows that the network coding maps for the case when $n_A \neq n_B$ can be obtained from network coding maps for the scenario in which the number of transmit antennas at A and B $n=\max \lbrace n_A, n_B \rbrace.$ Hence, in the rest of the paper it is assumed that $n_A=n_B=n$ and the network coding maps to be obtained are Latin Squares.

\begin{note}
If the maximum entry in the filled $n \times n$ Latin Square from which the Latin Rectangle is obtained is greater than $n-1,$ the obtained Latin Rectangle also can have entries greater than $n-1.$ Since the performance during the BC phase is dependent on the number of distinct entries in the filled Latin Rectangle, this adversely impacts the performance during the BC phase. In that case, obtaining the Latin Rectangle by direct completion of CPLR, instead of obtaining it from the Latin Square, may lead to a better performance.   
\end{note}
\subsection{Some Special Constructions of Latin Squares}
%In this subsection, it is assumed that the number of transmit antennas at the nodes A and B are equal, i.e., $n_A=n_B=n.$ The assumption is justified, since in the previous subsection it was shown that the Latin Rectangles which remove the singular fade subspaces for $n_A \times n_R \times n_B$ system can be obtained from the Latin Squares which remove the singular fade subspaces for $n \times n_R \times n$ system, where $n=\max\lbrace n_A,n_B\rbrace.$
%In this subsection, certain special constructions based on the structural properties of Latin Squares are provided. 

Recall that the rows and columns of the Latin Squares are indexed by vectors which belong to $\mathbb{Z}_M^{n}.$ By bit-wise XOR of two such vectors, it is meant the vector obtained by taking the bit-wise XOR of the individual components of the two vectors, after decimal to binary conversion. Every cell in the Latin Square corresponding to the bit-wise XOR mapping is filled with the decimal equivalent of the bit-wise XOR of the row index and the column index.

Consider the singular fade subspaces  $[\mathrm{span}([\Delta x_A \Delta x_B]^T)
]^\perp,$ which satisfy the condition that  $\Delta x_{A_i} = \pm \Delta x_{B_i}, \forall 1 \leq i \leq n.$ Let ${\mathcal{F}}_{\pm}$ denote the set of such singular fade subspaces. In the following lemma, it is shown that bit-wise XOR mapping removes all the singular fade subspaces which belong to ${\mathcal{F}}_{\pm}.$
%\begin{note}
%\end{note}
\begin{lemma}
\label{lemma_xor}
When the user nodes use $2^{\lambda}$-PSK constellations, the singular fade subspaces which belong to the set ${\mathcal{F}}_{\pm}$ are removed by bit-wise XOR mapping, for all $\lambda.$
\begin{proof}
By definition, for $f=\mathrm{span}([\Delta x_A \Delta x_B]^T)^\perp) \in {\mathcal{F}}_{\pm},$ the non-zero locations of $\Delta x_A$ and $\Delta x_B$ should match. Let $L$ be the number of non-zero components in $\Delta x_A$ and $\Delta x_B.$ Without loss of generality assume, $\Delta x_J=[\Delta x_{J_1} \; \Delta x_{J_2} \; \dotso \; \Delta x_{J_L} 0_{n-L}]^T,J \in \lbrace A,B \rbrace,$ (if the non-zero components of $\Delta x_A$ and $\Delta x_B$ appear in any other order, the indexing given for the transmit antennas can be permuted to get the assumed ordering). By definition, any $f \in \mathcal{F}_{\pm}$ should satisfy, $\Delta x_{A_i}= \pm \Delta x_{B_i}, 1 \leq i \leq L.$ Let $x_{A_i}=k_i, 1 \leq i \leq L$ and $x_{B_i}=l_i, 1 \leq i \leq L.$ For $\lbrace [k_1 \; k_2 \; \dotso k_L \; l_1 \; l_2 \: \dotso l_L]^T,[k'_1 \; k'_2 \; \dotso k'_L \; l'_1 \; l'_2 \: \dotso l'_L]^T \rbrace$ to be a singularity-removal constraint, since $\Delta x_{A_i}= \pm \Delta x_{B_i},$ it follows that,

\begin{equation}
\label{lemma_xor_eqn1}
e^{\frac{j k_i \pi}{M}}-e^{\frac{j k'_i \pi}{M}}= \pm \left(e^{\frac{j l_i \pi}{M}}-e^{\frac{j l'_i \pi}{M}}\right), 1 \leq i \leq L.
\end{equation}

Consider the case when  $\Delta x_{A_i}= \Delta x_{B_i}$.

Equating the magnitude and phase terms in \eqref{lemma_xor_eqn1} results in the following possibilities:
\begin{align}
\label{lemma_xor_eqn2}
&k_i+k'_i=l_i+l'_i+l_i,\\
\label{lemma_xor_eqn3}
&k_i-k'_i=l_i-l'_i \: \text{or} k_i-k'_i=l'_i-l_i+M.
\end{align}

Solving \eqref{lemma_xor_eqn2} and \eqref{lemma_xor_eqn3}, we get $k_i=l_i,k'_i=l'_i$ or $k_i=l'_i+\frac{M}{2},l_i=k'_i+\frac{M}{2}.$ 

Similarly for the case when $\Delta x_{A_i}= -\Delta x_{B_i},$ the conditions $k_i,l_i,k'_i$ and $l_i$ should satisfy are,
$k_i=l'_i,l_i=k'_i$ or $k_i=l_i+\frac{M}{2},k'_i=l'_i+\frac{M}{2}.$ Hence, it needs to be shown that the bit-wise XOR map places in the same cluster those vector pairs $(k,l)$ and $(k',l'),$ for which $k_i=l_i,k'_i=l'_i$ or $k_i=l'_i+\frac{M}{2},l_i=k'_i+\frac{M}{2},$ or $k_i=l'_i,l_i=k'_i$ or $k_i=l_i+\frac{M}{2},k'_i=l'_i+\frac{M}{2},$ or $ k_i=l'_i,l_i=k'_i$ or $k_i=l_i+\frac{M}{2},k'_i=l'_i+\frac{M}{2}, \forall 1 \leq i \leq L.$

The Latin Square corresponding to the bit-wise XOR map has the following properties: 
\begin{itemize}
\item
$k_i+k_i=0.$ Hence, it places vectors with $k_i=l_i,k'_i=l'_i$ in the same cluster. 
\item
$k_i+l_i=k_i-\frac{M}{2}+l_i-\frac{M}{2}.$ Hence, it places vectors with $k_i=l'_i+\frac{M}{2},l_i=k'_i+\frac{M}{2}$ in the same cluster.
\item
$k_i+l_i=l_i+k_i.$ Hence, it places vectors with $k_i=l'_i,l_i=k'_i$ in the same cluster.
\item
$k_i+k_i-\frac{M}{2}=l_i+l_i-\frac{M}{2}.$ Hence, it places vectors with $k_i=l_i+\frac{M}{2},k'_i=l'_i+\frac{M}{2}$ in the same cluster.
\end{itemize}

This completes the proof.
%See Appendix \ref{App_lemma_xor}.
\end{proof}
\end{lemma}

\begin{example}
Consider the $2 \times 2$ system with BPSK signal set. The singular fade subspaces 6-9 and 30-37 (12 in total) given in Table 1  belong to ${\mathcal{F}}_{\pm}$ and are removed by the bit-wise XOR map.
\end{example}

\begin{definition}
A Latin Square $L^T$ is said to be the Transpose of a Latin Square $L$, if $L^T(i,j)=L(j,i)$ for all $i,j \in \{0,1,2,..,M-1\}.$
\end{definition}

%%%%%%%%%%%%%%%%%%%%%%%%%%%%%%%%%%%%%%%%%%%%%%%

%Recall from Section II that if $\gamma e^{j\theta}$ is a singular fade state, then $\frac{1}{\gamma}e^{-j \theta}$  is also a singular fade state. The following Lemma shows that the transpose of the Latin Square that removes $\gamma e^{j\theta}$ removes the singular fade state $\frac{1}{\gamma}e^{-j \theta}$.
\begin{lemma}
\label{lemma_trans}
If the Latin Square $L$ removes the singular fade subspace $[\text{span} ([\Delta x_{A}^T \; \Delta x_{B}^T)]^T]^\perp,$ then the  Latin Square $L^T$ removes the singular fade subspace $[\text{span} ([\Delta x_{B}^T \; \Delta x_{A}^T)]^T]^\perp.$
\begin{proof}
The singular fade subspace $[\text{span} ([\Delta x_{A}^T \; \Delta x_{B}^T)]^T]^\perp$ can be viewed as the singular fade subspace $[\text{span} ([\Delta x_{B}^T \; \Delta x_{A}^T)]^T]^\perp$ with the users A and B interchanged. Interchanging the users is the same as interchanging the row and column indices, i.e., taking transpose.
\end{proof}
\end{lemma}

\begin{example}
For a $2  \times 2$ system with BPSK signal set, since the Latin Square given in Fig. \ref{lat_2} removes the singular fade subspace $[\mathrm{span}([0 \: {-2} \: {-2} \:\:2]^T)]^\perp,$ from Lemma \ref{lemma_trans}, its transpose shown in Fig. \ref{lat_3} removes the singular fade subspace $[\mathrm{span}([{-2} \:2 \:0 \: {-2} ]^T)]^\perp.$
\begin{figure}
\centering
\begin{tabular}{|c|c|c|c|c|}
\hline  & 00 & 01 & 10 & 11\\
\hline 00 & 3 & 1 & 0 & 2\\
\hline 01 & 0 & 2 & 1 & 3\\
\hline 10 & 2 & 0 & 3 & 1\\
\hline 11 & 1 & 3 & 2 & 0 \\
\hline
\end{tabular}
\caption{Latin Square that removes the singular fade subspace $[\mathrm{span}([{-2} \: 2 \: 0 \:\:\:{-2}]^T)]^\perp.$}
\label{lat_3}
\end{figure}
\end{example}
%%%%%%%%%%%%%%%%%%%%%%%%%%%%%%%%%%%%%%%%
\begin{definition} 
\cite{Sto} 
Two Latin Squares $L$ and $L$ $^{\prime}$ (using the same symbol set) are isotopic if there is a triple $(\textit{f,g,h}),$ where $f$ is a row permutation, $g$ is a column permutation and $h$ is a symbol permutation, such that applying these permutations on  $L$ gives $L^{\prime}.$
\end{definition}

Consider a vector $\Delta \tilde{x}=[ \Delta \tilde{x}_A^T \: \Delta \tilde{x}_B^T]^T,$ where $\Delta \tilde{x}_A$ and $\Delta \tilde{x}_B$ are obtained by the applying the  permutations  $\sigma_A$ and $\sigma_B$ on $\Delta {x}_A$ and $\Delta {x}_B$ respectively. Equivalently this can be viewed as applying the permutations $\sigma_A$ and $\sigma_B$ on the indices of the transmitting antennas at $n_A$ and $n_B$ respectively. Since the rows and columns of the Latin Squares which remove the singular fade subspaces are indexed by the vectors transmitted by nodes $A$ and $B$ respectively, applying the permutations $\sigma_A$ and $\sigma_B$ on the components of the row and column indices of the Latin Square which removes the singular fade subspace $[\text{span}(\Delta x)]^\perp$ results in an isotopic Latin Square which removes the singular fade subspace $[\text{span}(\Delta \tilde{x})]^\perp.$ This is stated as the following lemma. 

\begin{lemma}
\label{isotop1}
%Let $\tilde{x}_A \in \Delta \mathcal{S} ^ {n_A}$ and $\tilde{x}_B \in \Delta \mathcal{S} ^ {n_B}$ be obtained by applying the permutations $\sigma_A $ and $\sigma_B$ on $\Delta {x}_A\in \Delta \mathcal{S} ^ {n_A}$ and $\Delta {x}_B \in \Delta \mathcal{S} ^ {n_B}$ respectively. 
If a Latin Square $L$ removes the singular fade subspace $[\text{span}(\Delta x)]^\perp,$ the Latin Square $L'$ obtained by applying the permutation $\sigma_A$ on the components of the row indices and the permutation $\sigma_B$ on the components of the column indices of $L$ removes the singular fade subspace $[\text{span}(\Delta \tilde{x})]^\perp.$
\end{lemma}

\begin{example}
Consider the $2  \times 2$ system with BPSK signal set. As seen in Example 10, the XOR map given in Fig. \ref{lat_1} removes the singular fade subspaces 6-9 given in Table 1. Permuting the components of the row indices of the Latin Square in Fig. \ref{lat_1} (i.e., the first component becomes the second component and vice verse), results in the Latin Square Fig. \ref{lat_5}, which is the same as the Latin Square shown in Fig. \ref{lat_6}. From Lemma \ref{isotop1}, it follows that the Latin Square in Fig. \ref{lat_6} removes the singular fade subspaces 9-13 given in Table 1.
\begin{figure}
\centering
\begin{tabular}{|c|c|c|c|c|}
\hline  & 00 & 01 & 10 & 11\\
\hline 00 & 0 & 1 & 2 & 3\\
\hline 10 & 1 & 0 & 3 & 2\\
\hline 01 & 2 & 3 & 0 & 1\\
\hline 11 & 3 & 2 & 1 & 0 \\
\hline
\end{tabular}
\caption{Latin Square obtained from XOR map by the permutation of row indices}
\label{lat_5}
\end{figure}
\begin{figure}
\centering
\begin{tabular}{|c|c|c|c|c|}
\hline  & 00 & 01 & 10 & 11\\
\hline 00 & 0 & 1 & 2 & 3\\
\hline 01 & 2 & 3 & 0 & 1\\
\hline 10 & 1 & 0 & 3 & 2\\
\hline 11 & 3 & 2 & 1 & 0 \\
\hline
\end{tabular}
\caption{Latin Square which is same as the one shown in Fig. \ref{lat_5}}
\label{lat_6}
\end{figure}
\end{example}
%%%%%%%%%%%%%%%%
%\begin{lemma}
%\label{lemma_col_perm}
%Two Latin Squares $L$ and $L^\prime$ which remove the singular fade states $(\gamma, \theta)$ and $(\gamma, \theta^{\prime})$, respectively, (i.e., two singular fade states on %the same circle), are Isotopic that are obtainable one from another by a column permutation alone. If $\theta'-\theta = k\frac{2\pi}{M}$, $L'$ can be obtained by cyclic %shifting of the columns of $L$, $k$ times in the anticlockwise direction. 
%\end{lemma}

Consider the case when $M$-PSK signal set ($M$ any power of 2) is used at the end nodes. Consider two singular fade subspaces $[\text{span}(\Delta x)]^{\perp}$ and $[\text{span}(\Delta \bar{x})]^{\perp}$ which are such that the absolute values of the components of $\Delta x$ and $\Delta \bar{x}$ are equal, i.e., they differ only in the relative phase vectors. Let $[ {\phi_A ^T} \:  {\phi_B^T}]^T$ and $[ {{\phi'}_A ^T} \:  {{\phi'}_B^T}]^T$ be the relative phase vectors of $\Delta x$ and $\Delta \bar{x}$ respectively, where $\phi_A$ and $\phi'_A$ are of length $n-1$, and $\phi_B$ and $\phi'_B$ are of length $n$. Let $\Delta \phi_A=\phi_A-\phi'_A$  and $\Delta \phi_B=\phi_B-\phi'_B.$ Also let $\Delta \phi_{A_i} ={\frac{k_{A_i}2\pi}{M}}$ and $\Delta \phi_{B_i} ={\frac{k_{B_i}2\pi}{M}}.$ 

\begin{lemma}
\label{isotop2}
For $2^{\lambda}$-PSK signal set, let $L$ denote the Latin Square which removes the singular fade subspace $[\text{span}(\Delta x)]^{\perp}.$ The Latin Square $L'$ which removes the singular fade subspace $[\text{span}(\Delta \bar{x})]^{\perp}$ can be obtained from $L$ as follows: To the $i^{th}$ component of all the row indices of $L$ add $k_{A_i}, \forall 1 \leq i \leq n$ modulo $M$ and to the $i^{th}$ component of all the column indices of $L$ add $k_{B_i}, \forall 1 \leq i \leq n$ modulo $M,$ to obtain the Latin Square $L'.$ 
\begin{proof}
Rotating the phase of the $i^{th}$ component of $\Delta x_{A}$ by an angle $\frac{{{k_{A}}_i}2\pi}{M},$ can be viewed equivalently as rotating the phase of the signal set used by the antenna $n_i$ at node A, by the same angle. In the Latin Square $L$ that removes $[\text{span}(\Delta x)]^{\perp}$, this change can be effected by adding $k_{A_i}$ modulo $M$ to the $i^{th}$ component of the row indices. By a similar reasoning, $k_{B_i}$ needs to be added to the column indices of $L$ to obtain $L'.$ This completes the proof. 
\end{proof}
\end{lemma}

\begin{example}
Consider the $2  \times 2$ system with BPSK signal set. As discussed in Example 8, the Latin Square shown in Fig. \ref{lat_2} removes the singular fade subspace $[\text{span}([0 \: {-2} \:{-2} \:2]^T)]^\perp=[\text{span}([0 \: {2} \:{2} {-2}]^T)]^\perp.$ From Lemma \ref{isotop2}, adding $1$ modulo 2 to the second component of the row index in the Latin Square in Fig. \ref{lat_2}, we get the Latin Square shown in Fig. \ref{lat_7} which is the same as the one in Fig. \ref{lat_8}, which removes the singular fade subspace  $[\text{span}([0 \: {2} \:{-2}\: 2]^T)]^\perp.$
\begin{figure}
\centering
\begin{tabular}{|c|c|c|c|c|}
\hline  & 00 & 01 & 10 & 11\\
\hline 01 & 3 & 0 & 2 & 1\\
\hline 00 & 1 & 2 & 0 & 3\\
\hline 11 & 0 & 1 & 3 & 2\\
\hline 10 & 2 & 3 & 1 & 0 \\
\hline
\end{tabular}
\caption{Latin Square that removes the singular fade subspace $[\mathrm{span}([0 \: 2 \: -2 \:\:\:2]^T)]^\perp.$}
\label{lat_7}
\end{figure}
\begin{figure}
\centering
\begin{tabular}{|c|c|c|c|c|}
\hline  & 00 & 01 & 10 & 11\\
\hline 00 & 1 & 2 & 0 & 3\\
\hline 01 & 3 & 0 & 2 & 1\\
\hline 10 & 2 & 3 & 1 & 0 \\
\hline 11 & 0 & 1 & 3 & 2\\
\hline
\end{tabular}
\caption{Latin Square that is same as the Latin Square in Fig. \ref{lat_7}}
\label{lat_8}
\end{figure}
\end{example}

The usefulness of Lemmas \ref{lemma_trans}-\ref{isotop2} is that the set of all Latin Squares which remove all the singular fade subspaces can be obtained from a small set of Latin Squares. This is illustrated for the $2 \times 2$ system in the following subsection.

\subsection{Sufficient number of Latin Squares to be obtained for $2 \times 2$ system with $M$-PSK signal set}
Consider the $2 \times 2$ system with $M$-PSK ($M$ any power of 2) signal set used at A and B. For this case, the singular fade subspaces are of the form $[\text{span}(\Delta x)]^\perp$ where $\Delta x=[\Delta x_{A_1}\: \Delta x_{A_2} \:\Delta x_{B_1} \: \Delta x_{B_2}]^T \in \Delta \mathcal{S}^4.$  Let $k$ be the number of non-zero components in $\Delta x.$  The following lemma gives the sufficient number of Latin Squares from which the Latin Squares which remove all the removable singular fade subspaces can be obtained.\\

\begin{lemma}
For the $2 \times 2$ system, with $2^{\lambda}$-PSK signal set, the sufficient number of Latin Squares from which the Latin Squares which remove all the removable singular fade subspaces of the form $[\text{span}(\Delta x)]^\perp,$ with $k$ non-zero components in $\Delta x$ can be obtained are given by,

\vspace{-0.5 cm}
{\centering
$$\begin{array}{|c|c|}
\hline k & \text{No.of Latin Squares}\\
\hline 2   & \frac{M^2}{8}-\frac{M}{4}+1\\
\hline 3   & \frac{M^3}{16}+\frac{M^2}{8}-\frac{M}{2}+1  \\
\hline 4   & \frac{M^4}{128}+\frac{M^3}{32}-\frac{11M^2}{32}-\frac{3M}{8}+1.\\
\hline
\end{array}$$
}
\begin{proof}
\noindent \textit{Case 1: k=4}

For this case, from Lemma \ref{sing_non_rem}, there are $\left [\left(\frac{M}{2}\right)^4-\frac{M}{2}+1\right]M^3$ singular fade subspaces.

For vectors $\Delta x$ and $\Delta \tilde{x}$ such that $\vert \Delta x_i\vert=\vert \Delta \tilde{x}_i \vert, \forall 1 \leq i \leq n$ i.e., $\Delta x$ and $\Delta \tilde{x}$ differ only in their relative phase vectors, from Lemma \ref{isotop1}, the Latin Square which removes $[\text{span}(\Delta \tilde{x})]^\perp$ can be obtained by row and column permutations. This reduces the total number of Latin Squares to be obtained by a factor $M^3.$ In other words the total number of Latin Squares to be obtained is upper bounded by $\left [\left(\frac{M}{2}\right)^4-\frac{M}{2}+1\right].$ 

Consider the set $\Delta S_{\setminus 0}^2 =\lbrace \lbrace \Delta x_1 , \Delta x_2 \rbrace \in \Delta \mathcal{S} ^2 :  \Delta x_1 , \Delta x_2 \neq 0\rbrace.$ There are ${\frac{M}{2} \choose 2}+ {\frac{M}{2} \choose 1}=\frac{M(M+2)}{8}$ choices for the pair $\lbrace \Delta x_1 , \Delta x_2 \rbrace.$ If $\Delta x$ and $\Delta \tilde{x}$ such that the unordered pairs $\lbrace \Delta{x_A} \Delta{x_B} \rbrace$ and $\lbrace \Delta{\bar{x}_A} \Delta{\bar{x}_B} \rbrace$ are the same, from Lemma \ref{isotop2}, the Latin Square which removes the singular fade subspace $[\text{span}(\Delta \bar{x})]^\perp$ can be obtained from the one which removes $[\text{span}(\Delta {x})]^\perp.$ 

Hence the number of Latin Squares to be obtained is equal to the number of ways of choosing two elements from the set $\Delta S_{\setminus 0}^2$ minus the number of cases which results in the same singular fade subspaces. The total number of such possibilities is, 

\begin{align*}
&{\frac{M(M+2)}{8} \choose 2}+{\frac{M(M+2)}{8} \choose 1}-\frac{M}{2}+1=\\
&\hspace{3 cm}\frac{M^4}{128}+\frac{M^3}{32}-\frac{11M^2}{32}-\frac{3M}{8}+1.
\end{align*}

\noindent \textit{Case 2: k=3}
For this case, from Lemma \ref{sing_non_rem}, the number of removable singular fade subspaces are $\left [\left(\frac{M}{2}\right)^3-\frac{M}{2}+1\right]M^2.$ Considering only those singular fade subspaces for which $\vert \Delta x \vert$ is distinct, the total number of Latin Squares to be obtained for this case is less than  $\left [\left(\frac{M}{2}\right)^3-\frac{M}{2}+1\right].$

The number of Latin Squares to be obtained is equal to the number of ways of choosing an element from the set $\Delta S_{\setminus 0}^2$ and an element from the set $\Delta \mathcal{S} \setminus \lbrace 0 \rbrace$ minus the number of cases which results in the same singular fade subspaces. The total number of such possibilities is, $$\left[{\frac{M(M+2)}{8}}\right]\frac{M}{2}-\frac{M}{2}+1=\frac{M^3}{16}+\frac{M^2}{8}-\frac{M}{2}+1.$$

Case 3: $k=2$

For this case, from Lemma \ref{sing_non_rem}, there are $[\frac{M^2}{4}-\frac{M}{2}+1]M^3$ singular fade subspaces. The total number of Latin Squares to be obtained is equal to the total number of ways of choosing two elements from  $\Delta \mathcal{S} \setminus \lbrace 0 \rbrace$ minus the number of cases which result in the same singular fade subspaces. The total number of such possibilities is, $$ {\frac{M}{2} \choose 2} +{\frac{M}{2} \choose 1} - \frac{M}{2}+1=\frac{M^2}{8}-\frac{M}{4}+1.$$
\end{proof}
\label{LS_set}
\end{lemma}

\begin{note}
 Lemma \ref{LS_set} provides only the sufficient number of Latin Squares from which all the Latin Squares can be obtained. The actual number can be lesser than the number given in Lemma \ref{LS_set}, since the same Latin Square can remove more than one singular fade subspace. For example, from Lemma \ref{lemma_xor} it follows that the bit-wise XOR map removes more than one singular fade subspace. 
\end{note}

\section{ILLUSTRATIONS FOR THE $2 \times 2$ SYSTEM WITH QPSK SIGNAL SET}

Consider the $2 \times 2$ system with QPSK signal set. The non-zero points of the difference constellation of the QPSK signal set $\Delta \mathcal{S}$ lie on two circle with radii $\sqrt{2}$ and $2.$ The four points on the circles with radius $\sqrt{2}$ have phase angles $\frac{2m\pi}{M}$ and the 4 points on the circle with radius $2$ have phase angles $\frac{(2m+1)\pi}{M}, 0 \leq m \leq 3.$ 

  As seen in Example 7, there are 1456 removable singular fade subspaces, out of which 960 have $k=4.$ From Lemma \ref{LS_set}, it follows that the set of Latin Squares which remove all these 960 singular fade subspaces can be obtained from 5 Latin Squares, which remove the following singular fade subspaces: 
\begin{align*}
&f_1=[\text{span}([{\sqrt{2}+j\sqrt{2}}\:{\sqrt{2}+j\sqrt{2}}\:{\sqrt{2}+j\sqrt{2}}\:{\sqrt{2}+j\sqrt{2}}])]^\perp,\\
 &f_2= [\text{span}([{\sqrt{2}+j\sqrt{2}}\:\sqrt{2}\:{\sqrt{2}+j\sqrt{2}}\:{\sqrt{2}+j\sqrt{2}}])]^\perp,\\
 &f_3=[\text{span}([{\sqrt{2}+j\sqrt{2}}\:\sqrt{2}\:\sqrt{2}\:\sqrt{2}])]^\perp,\\
 &f_4=[\text{span}([{\sqrt{2}+j\sqrt{2}}\:\sqrt{2}\:{\sqrt{2}+j\sqrt{2}}\:\sqrt{2}])]^\perp \text{ and}\\
 &f_5=[\text{span}([{\sqrt{2}+j\sqrt{2}}\:{\sqrt{2}+j\sqrt{2}}\:\sqrt{2}\:\sqrt{2}])]^\perp.
 \end{align*}
  From Lemma \ref{lemma_xor}, the singular fade subspaces $f_1$ and $f_4$ are removed by the bitwise-XOR map, given in Fig. \ref{LS_1_4}, given in Appendix B. The Latin squares which remove the other three singular fade subspaces $f_2,f_3$ and $f_5$ are given in Fig. \ref{LS_2}-\ref{LS_5} in Appendix B. 

For $k=3,$ the Latin Squares which remove all the 448 singular fade subspaces can be obtained from the Latin Squares given in Fig. \ref{LS_6}-\ref{LS_10} in Appendix B, which remove the following 5 singular fade subspaces: 
\begin{align*}
&f_6=[\text{span}([0\:{\sqrt{2}+j\sqrt{2}}\:{\sqrt{2}+j\sqrt{2}}\:{\sqrt{2}+j\sqrt{2}}]^T)]^{\perp},\\
&f_7=[\text{span}([0\:{\sqrt{2}+j\sqrt{2}}\:\sqrt{2}\:{\sqrt{2}+j\sqrt{2}}]^T)]^{\perp},\\
&f_8=[\text{span}([0\:{\sqrt{2}+j\sqrt{2}}\:\sqrt{2}\:\sqrt{2}]^T)]^{\perp},\\
&f_9=[\text{span}([0\:{\sqrt{2}+j\sqrt{2}}\:{\sqrt{2}+j\sqrt{2}}\:\sqrt{2}]^T)]^{\perp} \text{ and}\\ &f_{10}=[\text{span}([0\:\sqrt{2}\:\sqrt{2}+j\sqrt{2}\:\sqrt{2}]^T)]^{\perp}.
\end{align*}
For $k=2$, there are 48 removable singular fade subspaces. From Lemma \ref{LS_set}, the Latin Squares which remove all these 48 singular fade subspaces can be obtained from the 2 Latin Squares which remove the singular fade subspaces $f_{11}=[\text{span}([0\:\sqrt{2}\:{0}\:{\sqrt{2}}]^T)]^{\perp}$ and $f_{12}=[\text{span}([0\:{\sqrt{2}+j\sqrt{2}}\:{0}\:\sqrt{2}]^T)]^{\perp}$. From Lemma \ref{lemma_xor}, it follows that the XOR map given in Fig. \ref{LS_1_4} in Appendix B removes $f_{11}.$ The Latin Square that removes $f_{12}$ is given in Fig. \ref{LS_12} in Appendix B. To sum up, the Latin Squares which remove all the 1496 removable singular fade subspaces, are obtainable from the 10 Latin Squares given in Fig. \ref{LS_set1} in Appendix B.

\section{OBTAINING LATIN SQUARES FOR THE $n \times n$ SYSTEM FROM LATIN SQUARES OF LOWER ORDER}

In this section, it is shown that most of the Latin Squares which remove the singular fade subspaces of the $n \times n$ system, $n \geq 2$ are obtainable from the Latin Squares which remove the singular fade subspaces of $m \times m$ system, where $m < n.$

\begin{definition}
For two vectors $y$ and $z$ of length $2a$ and $2b$ respectively, the compound vector of $y$ and $z,$ denoted as $\mathrm{comp}(y,z)$ is the vector of length $2a+2b$ given by,
$$[y_{[1:a]}^T  \quad z_{[1:b]}^T\quad y_{[a+1:2a]}^T \quad z_{[b+1:2b]}^T]^T.$$
\end{definition}
For example, for two vectors $y=[1 \; 2\; 3\; 4]^T$ and $z=[0 \;5]^T$, $\mathrm{comp}(y,z)=[1 \;2 \;0 \;3 \;4 \;5]^T.$

For a Latin Square $L,$ let $L_{[i:j,k:l]}$ denote the $(j-i+1) \times (l-k+1)$ array  obtained by taking only the $i^{th}$ to $j^{th}$ rows and $k^{th}$ to $l^{th}$ columns of $L.$ Let $L+c$ denote the Latin Square obtained by adding integer $c$ to all the cells of $L.$ Let $\max \lbrace L \rbrace$ denote the maximum among all the integers filled in the cells of the Latin Square $L$. 
\begin{definition}
The Cartesian product of the two Latin Squares $L$ of order $M^a$ and $\bar{L}$   of order $M^b,$ denoted as $(L \otimes \bar{L}),$ is the Latin Square of order $M^{a+b}$ for which 

{\vspace{-.3 cm}
\footnotesize $$(L \otimes \bar{L})_{\left[\left(i-1\right)M^a+1:iM^a,\left(j-1\right)M^a+1:jM^a\right]}=L+\bar{L}{\left(i,j\right)}\left(\max\lbrace {L} \rbrace+1\right),$$} where $1 \leq i,j \leq M^b.$ 
\end{definition}

\begin{figure}
\centering
\subfigure[The Latin Squares $L$ and $\bar{L}$]{
\includegraphics[height=1.5 in,width=1 in]{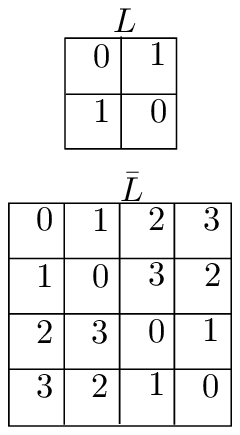}
\label{fig:cp_ex1} 
}
\subfigure[The Latin Square $L \otimes \bar{L}$]{
\includegraphics[height=2 in,width=2 in]{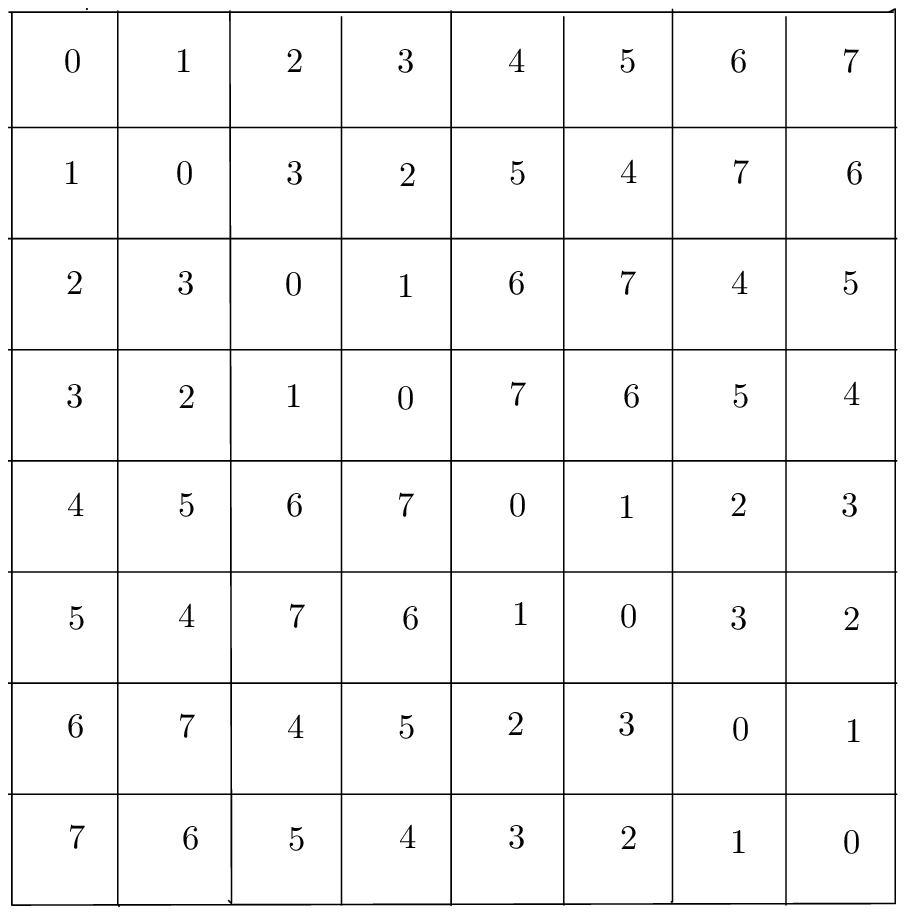}
\label{fig:cp_ex2} 
}
\caption{Example illustrating the notion of Cartesian product of Latin Squares}
\label{fig:cp_ex} 
\end{figure}
For example, the Cartesian product of the Latin Squares $L$ and $\bar{L}$ of order $2^1$ and $2^2$ respectively shown in Fig. \ref{fig:cp_ex1}, is the Latin Square of order $2^3$ shown in Fig. \ref{fig:cp_ex2}.
 
Consider two vectors $\Delta x \in \Delta \mathcal{S}^{2a}$ and $\Delta \bar{x} \in \Delta \mathcal{S}^{2b}.$ The vector subspaces $[\text{span}(\Delta x)]^\perp$ and  $[\text{span}(\Delta \bar{x})]^\perp$ are singular fade subspaces for the $a \times a$ and $b \times b$ systems respectively. The following lemma shows that the Latin Square which removes all the singular fade subspaces of the form $[\text{span}(\mathrm{comp}(\Delta x, k\Delta \bar{x}))]^\perp, k \in \mathbb{C}$ can be obtained by taking the Cartesian product of the Latin Squares which remove $[\text{span}(\Delta x)]^\perp$ and $[\text{span}(\Delta \bar{x})]^\perp.$  

\begin{note}
For $\bar{x} \in \Delta \mathcal{S}^{2b}$ and $k \in \mathbb{C}$ such that $k\Delta \bar{x} \in \mathcal{S}^{2b},$ the vector subspaces of $\mathbb{C}^{2b}$ $[\text{span}(\Delta \bar{x})]^\perp$ and $[\text{span}(k\Delta \bar{x})]^\perp$ are the same. But the vector subspaces of $\mathbb{C}^{2a+2b}$ $[\text{span}(\mathrm{comp}(\Delta x, \Delta \bar{x}))]^\perp$ and $[\text{span}(\mathrm{comp}(\Delta x, k\Delta \bar{x}))]^\perp, k \neq 1$ are different.
\end{note}
\begin{lemma}
Let $L$ and $\bar{L}$ respectively denote the Latin Squares of order $M^a \times M^a$ and $M^b \times M^b,$ which remove the singular fade subspaces $[\text{span}(\Delta x)]^\perp$ of the $a \times a$ system, and  $[\text{span}(\Delta \bar{x})]^\perp$ of the $b \times b$ system. The Latin Square $L \otimes \bar{L}$ removes all the singular fade subspaces of the form $[\text{span}(\mathrm{comp}(\Delta x, k\Delta \bar{x}))]^\perp, k \in \mathbb{C}.$
\begin{proof}
For the $a \times a$ system, let

 {\footnotesize 
\begin{align*} 
 &\left \lbrace \left(\left[x_{1_A} \: x_{2_A} \dotso x_{a_{A}}\right],\left[x_{1_B} \: x_{2_A} \dotso x_{a_{B}}\right]\right),\right.\\
&\hspace{2.5 cm} \left.\left(\left[x'_{1_A} \: x'_{2_A} \dotso x'_{a_{A}}\right],\left[x'_{1_B} \: x'_{2_A} \dotso x'_{a_{B}}\right]\right)\right \rbrace
 \end{align*}}denote a singularity removal constraint for the singular fade subspace $[\text{span}(\Delta x)]^\perp.$
 Similarly, for the $b \times b$ system, let

 {\footnotesize 
\begin{align*} 
 &\left \lbrace \left(\left[y_{1_A} \: y_{2_A} \dotso y_{b_{A}}\right],\left[y_{1_B} \: y_{2_A} \dotso y_{b_{B}}\right]\right),\right.\\
&\hspace{2.5 cm} \left.\left(\left[y'_{1_A} \: y'_{2_A} \dotso y'_{b_{A}}\right],\left[y'_{1_B} \: y'_{2_A} \dotso y'_{b_{B}}\right]\right)\right \rbrace
 \end{align*}}denote a singularity removal constraint for the singular fade subspace $[\text{span}(\Delta \bar{x})]^\perp.$ Since $\text{span}(\Delta \bar{x})=k\text{span}(\Delta \bar{x}),$ 

 {\footnotesize 
\begin{align*} 
 &\left \lbrace \left(\left[ky_{1_A} \: ky_{2_A} \dotso ky_{b_{A}}\right],\left[ky_{1_B} \: ky_{2_A} \dotso ky_{b_{B}}\right]\right),\right.\\
&\hspace{2.5 cm} \left.\left(\left[ky'_{1_A} \: ky'_{2_A} \dotso ky'_{b_{A}}\right],\left[ky'_{1_B} \: ky'_{2_A} \dotso ky'_{b_{B}}\right]\right)\right \rbrace
 \end{align*}}is also a singularity removal constraint for the singular fade subspace $[\text{span}(\Delta \bar{x})]^\perp.$
 
It can be verified that the Cartesian product of the Latin Squares which remove the singular fade subspaces  $[\text{span}(\Delta {x})]^\perp$ and $[\text{span}(\Delta \bar{x})]^\perp$ satisfy the Cartesian product of their constraints. In other words, the Latin Square $(L \otimes \bar{L})$ satisfies the constraint, 

{\footnotesize 
\begin{align*} 
 &\left \lbrace \left(\left[x_{1_A} \: x_{2_A} \dotso x_{a_{A}}\: ky_{1_A} \: ky_{2_A} \dotso ky_{b_{A}}\right],\right.\right.\\
&\hspace{2.5 cm}\left.\left[x_{1_B} \: x_{2_A} \dotso x_{a_{B}} \: ky_{1_B} \: ky_{2_A} \dotso ky_{b_{B}}\right]\right),\\
&\hspace{0.2 cm}\left(\left[x'_{1_A} \: x'_{2_A} \dotso x'_{a_{A}}\: ky'_{1_A} \: ky'_{2_A} \dotso ky'_{b_{A}}\right],\right.\\
&\hspace{2.5 cm}\left.\left.\left[x'_{1_B} \: x'_{2_A} \dotso x'_{a_{B}} \: ky'_{1_B} \: ky'_{2_A} \dotso ky'_{b_{B}}\right]\right)\right \rbrace,
 \end{align*}}which is a singularity removal constraint for the singular fade subspace $[\text{span}(\mathrm{comp}(\Delta x, k\Delta \bar{x}))]^\perp$ of the $(a+b) \times (a+b)$ system. This completes the proof.
\end{proof}
\label{lemma_LS_from_lower}
\end{lemma}

From Lemma \ref{lemma_LS_from_lower}, it follows that Latin Squares for those singular fade subspaces for the $n \times n$ system,  which are expressible as $[\text{span}(\mathrm{comp}(\Delta x, k\Delta \bar{x}))]^\perp,$ where $[\text{span}(\Delta x)]^\perp$ and $[\text{span}(\Delta \bar{x})]^\perp$ are removable singular fade subspaces for $a \times a$ and $b \times b$ systems respectively, for some choices of $a<n$ and $b<n$ such that $a+b=n,$ are obtainable from the Latin Squares for the $a \times a$ and $b \times b$ systems.

\begin{example}
Consider the $3 \times 3$ system with BPSK signal set. Consider the singular fade subspace $[\text{span}([2 \:-2 \: 2 \: 2 \:2 \:-2])]^\perp.$ The Latin Square $L$ shown in Fig. \ref{fig:cp_ex1} removes the singular fade subspace $[\text{span}([2 \; 2]^T)^\perp$ of the $1 \times 1$ system and the Latin Square $\bar{L}$ shown in Fig. \ref{fig:cp_ex1} removes the singular fade subspace $[\text{span}([-2 \: 2  \:2 \:-2])]^\perp$ of the $2 \times 2$ system. Since $[2 \:-2 \: 2 \: 2 \:2 \:-2]=\text{comp}([2\: 2],[-2 \: 2  \:2 \:-2]),$ the Latin Square $L \otimes \bar{L}$ shown in Fig. \ref{fig:cp_ex2} removes the singular fade subspace $[\text{span}([2 \:-2 \: 2 \: 2 \:2 \:-2])]^\perp$ of the $3 \times 3$ system.
\end{example}

 From Lemma \ref{sing_non_rem} it follows that the number of non-removable singular fade subspaces for the $n \times n$ system with $M$-PSK signal set is of order $M^{4n-1}$ and the singular fade subspaces of the form $[\text{span}(\Delta x)]^\perp,$ where $\Delta x$ has all the components to be non-zero contributes to the maximum number of singular fade states which is of order $M^{4n-1}$. It is easy to verify that a vector $\Delta x$ with all the components to be non-zero can always be written of the form $\text{comp}(\Delta x_1, \Delta x_2),$ where $\Delta x_1$ and $\Delta x_2$ are of lengths $a$ and $b$, where $a,b<n$ and $a+b=n,$ such that $[\text{span}(\Delta x_1)]^\perp$ and $[\text{span}(\Delta x_2)]^\perp$ are removable singular fade subspaces for the $a \times a$ and $b \times b$ systems respectively. Hence, it follows that most of the Latin Squares which remove the singular fade subspaces for the $n \times n$ system are obtainable from Latin Squares of lower order.
%The following lemma gives the exact number of Latin Squares for the $n \times n$ system which are obtainable from Latin Squares of lower order.  
\section{SIMULATION RESULTS}
  The proposed scheme is based on the removal of all the singular fade subspaces, i.e., a minimum cluster distance greater than zero is ensured for all realizations of the channel fade coefficient matrices. Also, it is ensured that the number of clusters in the clustering, which is the same as the size of the signal set used during the BC phase is minimized. As a reference scheme, we consider the case when bit-wise XOR network code is used at R, irrespective of the channel conditions. Since XOR network does not remove all the singular fade subspaces, the proposed scheme is expected to perform better than the pure XOR network code based scheme, which is confirmed by the simulation results. All the simulation results presented are for the case when QPSK signal set is used at the end nodes. The number of antennas at all the three nodes are taken to be two.

\begin{figure}[htbp]
\centering
\vspace{-.2 cm}
\includegraphics[totalheight=2.75in,width=3.75in]{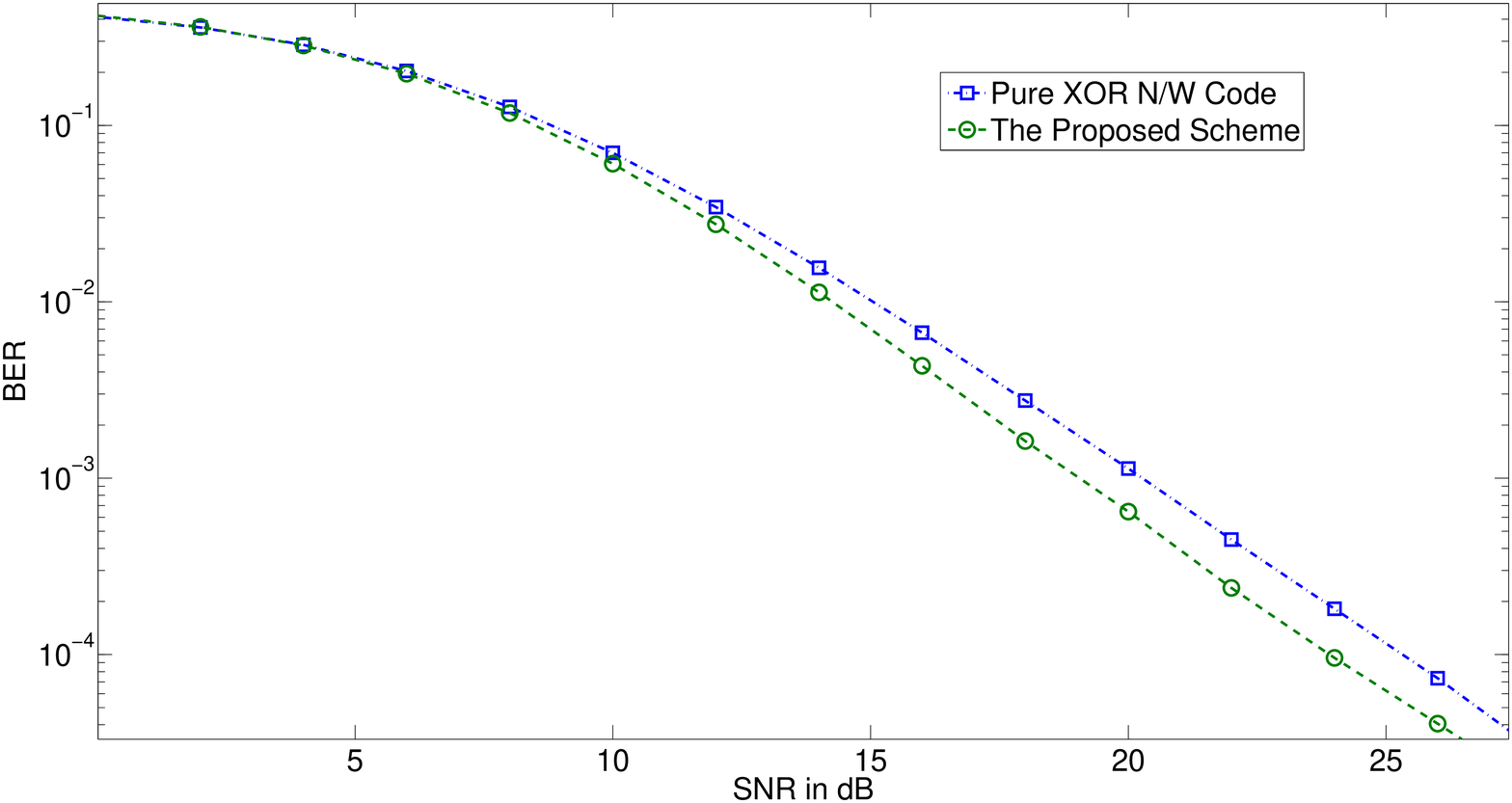}
\vspace{-.75 cm}
\caption{SNR vs BER for different schemes for QPSK signal set for a Rayleigh fading scenario with channel variances 0 dB}	
\label{fig:ber_rayleigh_0db}	
\end{figure}
%
%\begin{figure}[htbp]
%\centering
%\vspace{-.5 cm}
%\includegraphics[totalheight=2.75in,width=3.75in]{tput_rayleigh_0db_MIMO}
%\vspace{-.75 cm}
%\caption{SNR vs Throughput for different schemes for QPSK signal set for a Rayleigh fading scenario with channel variances 0 dB}	
%\label{fig:tput_rayleigh_0dB_MIMO}	
%\end{figure}
%%
%\begin{figure}[htbp]
%\centering
%\vspace{-.5 cm}
%\includegraphics[totalheight=2.75in,width=3.75in]{ber_a_rayleigh_5db}
%\vspace{-.75 cm}
%\caption{SNR vs BER for different schemes at node A for 8-PSK signal set for a Rayleigh fading scenario with channel power ratio 5 dB}	
%\label{fig:ber_a_rayleigh_5db}	
%\end{figure}
%
\begin{figure}[htbp]
\centering
\vspace{-.5 cm}
\includegraphics[totalheight=2.75in,width=3.75in]{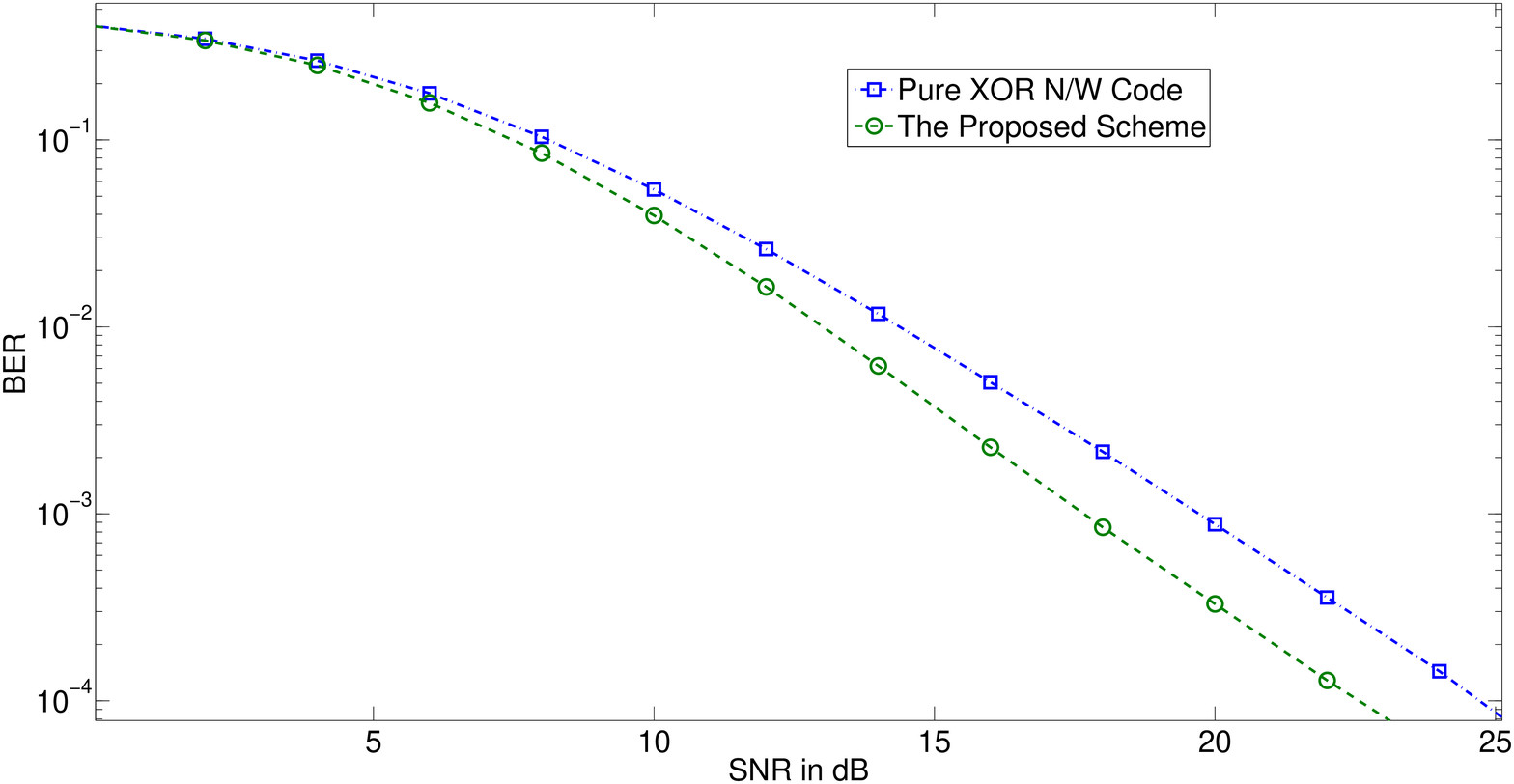}
\vspace{-.75 cm}
\caption{SNR vs BER for different schemes at node B for QPSK signal set for a Rician fading scenario with a Rician factor of 10 dB}	
\label{fig:ber_rician}	
\end{figure}
%%
%\begin{figure}[htbp]
%\centering
%\vspace{-.5 cm}
%\includegraphics[totalheight=2.75in,width=3.75in]{tput_rician}
%\vspace{-.75 cm}
%\caption{SNR vs throughput for different schemes for QPSK signal set for a Rician fading scenario with a Rician factor of 10 dB}	
%\label{fig:tput_rician}	
%\end{figure}

  The additive noises at the nodes are taken to be of unit variance. It is assumed that signal sets of equal energies are used at all the nodes, which is defined to be the Signal to Noise Ratio (SNR). Fig. \ref{fig:ber_rayleigh_0db} shows the SNR vs BER performance for the proposed scheme and the for the case when XOR network code is used irrespective of the channel condition, for the case when each one of the entries of the matrices $H_A$, $H_B$, $H'_A$ and $H'_B$ follow i.i.d. Rayleigh distribution, with unit variance. Fig. \ref{fig:ber_rician} shows a similar plot for the case when each one of the entries of the matrices $H_A$, $H_B$, $H'_A$ and $H'_B$ follow i.i.d. Rician distribution with a Rician Factor\footnote{Rician factor is the power ratio between the line of sight and scattered components.} of 10 dB and unit variances. From Fig. \ref{fig:ber_rayleigh_0db} and Fig. \ref{fig:ber_rician}, it can be seen that the diversity order is two, for the proposed scheme as well as the scheme which uses the conventional XOR network code. Also, it can be seen that for the Rayleigh and Rician fading scenarios considered, at a BER of $10^{-4},$ the proposed scheme provides a gain of 1.5 dB and 2.5 dB respectively over the conventional XOR network code based scheme. 

\section{Discussion} 
An adaptive network coding scheme using Latin Rectangles for the wireless two-way  relaying scenario was proposed. The scheme was based on the removal of a finite number of vector subspaces referred to as the removable singular fade subspaces by proper choice of network coding maps. For a particular channel realization, among all the network coding maps which remove the singular fade subspaces, the one which maximizes the minimum clustering distance is chosen. In this way, the set of all possible channel realizations, which is $\mathbb{C}^{n_A+n_B}$ for the $n_A \times n_B$ system, is quantized in  to a finite number of regions, based on which one of the network coding maps obtained optimizes the performance. Such a quantization was obtained analytically in \cite{VNR} for the $1 \times 1$ system. Obtaining the quantization for a general $n_A \times n_B$ system is an interesting problem for further investigation.  
\section*{Acknowledgement}
This work was supported  partly by the DRDO-IISc program on Advanced Research in Mathematical Engineering through a research grant as well as the INAE Chair Professorship grant to B.~S.~Rajan.

%%%%%%%%%%%%%%%%%%%%%%%%%%%%%%%%%%%%%%%%%%%%%%%%%%%%%%%%%%%%%%

\begin{appendices}
\section{PROOF OF LEMMA \ref{sing_non_rem}}
\subsection*{Number of non-removable singular fade subspaces:}
 Consider the case when the non-removable singular fade subspaces are of the form $[\mathrm{span}([\Delta x_A \; 0_{n_B}]^T)]^\perp.$ Let $k$ be the number of non-zero components of $\Delta x_A.$  Assume that the relative phase vector of $\Delta x_A$ is fixed. Since the points in $\Delta \mathcal{S}$ lie on $\frac{M}{2}$ circles, there are $(\frac{M}{2})^{k}$ possibilities for the absolute values of the non-zero components of $\Delta x_A.$ But out of these, if the absolute values of all the components of $\Delta x_A,$ are equal, from Lemma \ref{lemma_same_space}, the resulting singular fade subspaces are the same. Subtracting all such cases which are $\frac{M}{2}$ in number and adding one to account for all such cases, results in $\left[\left(\frac{M}{2}\right)^{k}-\frac{M}{2}+1 \right]$ singular fade subspaces. From Lemma \ref{lemma_same_space}, given the same absolute values of the non-zero components of $\Delta x_A,$ distinct singular fade subspaces results for distinct relative phase vectors of  $\Delta x_A.$ Since there are $M^{k-1}$ distinct possibilities for the relative phase vector, there are $\left[\left(\frac{M}{2}\right)^{k}-\frac{M}{2}+1 \right]M^{k-1}$ singular fade subspaces for the case considered with $k$ non-zero components. Summing over all possible values of $k,$ we have $\sum_{k=1}^{n_A} {n_A \choose k}  \left[\left(\frac{M}{2}\right)^{k}-\frac{M}{2}+1 \right]M^{k-1}$ singular fade subspaces for the case considered. Similarly, for the case when the non-removable singular fade subspaces are of the form $[\mathrm{span}([0_{n_A} \; \Delta x_B]^T)]^\perp,$ there are $\sum_{l=1}^{n_B} {n_B \choose l}  \left[\left(\frac{M}{2}\right)^{l}-\frac{M}{2}+1 \right]M^{l-1}$ singular fade subspaces.  

\subsection*{Number of removable singular fade subspaces:}
Let $k$ be the number of non-zero entries in $\Delta x=[\Delta x_A^T \; \Delta x_B^T]^T.$ Since $\Delta x_A \neq 0_{n_A}$ and $\Delta x_B \neq 0_{n_B},$ $k$ should be at least 2. Without loss of generality, assume that $n_B \geq n_A.$ 

\noindent
\textit{Case 1:} $2 \leq k \leq n_A$ 

The $k$ non-zero components of $\Delta x$ can be chosen in $n_A+n_B \choose k$ ways. But this also includes the cases when all the $k$ non-zero components occur in $\Delta x_A$ and $\Delta x_B=0_{n_B}$ or vice verse. Subtracting out those cases, the $k$ non-zero components of $\Delta x$ can be chosen in ${n_A+n_B \choose k}-{n_A \choose k}- {n_B \choose k}$ ways. For each one of those possibilities, following an approach similar to the one used in the proof of Lemma \ref{sing_non_rem}, it can be shown that there are $\left[(\frac{M}{2})^k-\frac{M}{2}+1\right]M^{k-1}$ singular fade subspaces. Hence, for this case there are $\left[{n_A+n_B \choose k}-{n_A \choose k}- {n_B \choose k}\right]\left[(\frac{M}{2})^k-\frac{M}{2}+1\right]M^{k-1}$ possible  singular fade subspaces in total. 

\noindent\textit{Case 2:} $n_A < k \leq n_B$

The $k$ non-zero components of $\Delta x$ can be chosen in ${n_A+n_B \choose k}-{n_B \choose k}$ ways, since the case when all the $k$ non-zero components occur in $\Delta x_B,\mathrm{ i.e.,} \: \Delta x_A=0_{n_A},$ needs to be excluded. Hence, there are $\left[{n_A+n_B \choose k}- {n_B \choose k}\right]\left[(\frac{M}{2})^k-\frac{M}{2}+1\right]M^{k-1}$ possibilities in total for this case.

\noindent\textit{Case 3:} $n_B < k \leq n_A+n_B$

The $k$ non-zero components of $\Delta x$ can be chosen in ${n_A+n_B \choose k}$ ways. This results in ${n_A+n_B \choose k}\left[(\frac{M}{2})^k-\frac{M}{2}+1\right]M^{k-1}$ singular fade subspaces in total for this case. 

For $2 \leq k \leq n_A+n_B,$ summing up the number of possibilities obtained in the above three cases and defining $a \choose b$ to be zero for $b>a,$ the number of removable singular fade subspaces is as given in the statement of the lemma. This completes the proof.
\section{LATIN SQUARES FOR THE 2 $\times$ 2 SYSTEM WITH 4-PSK SIGNAL SET}
For the $2 \times 2$ system with 4-PSK signal set, the Latin Squares from which the set of Latin Squares which remove all the removable singular fade subspaces are obtainable,  are given in the next three pages.
\newpage 
\begin{figure*}
\centering
\subfigure[Latin Square that removes the singular fade subspaces $f_1,$ $f_4$ and $f_{11}$]{
\begin{footnotesize}\begin{tabular}{|l|c|c|c|c|c|c|c|c|c|c|c|c|c|c|c|c|}
\hline
&\textbf{0 0}&\textbf{0 1}&\textbf{0 2}&\textbf{0 3}&\textbf{1 0}&\textbf{1 1}&\textbf{1 2}&\textbf{1 3}&\textbf{2 0}&\textbf{2 1}&\textbf{2 2}&\textbf{2 3}&\textbf{3 0}&\textbf{3 1}&\textbf{3 2}&\textbf{3 3}\\\hline
\textbf{0 0}&0&1&2&3&4&5&6&7&8&9&10&11&12&13&14&15\\\hline
\textbf{0 1}&1&0&3&2&5&4&7&6&9&8&11&10&13&12&15&14\\\hline
\textbf{0 2}&2&3&0&1&6&7&4&5&10&11&8&9&14&15&12&13\\\hline
\textbf{0 3}&3&2&1&0&7&6&5&4&11&10&9&8&15&14&13&12\\\hline
\textbf{1 0}&4&5&6&7&0&1&2&3&12&13&14&15&8&9&10&11\\\hline
\textbf{1 1}&5&4&7&6&1&0&3&2&13&12&15&14&9&8&11&10\\\hline
\textbf{1 2}&6&7&4&5&2&3&0&1&14&15&12&13&10&11&8&9\\\hline
\textbf{1 3}&7&6&5&4&3&2&1&0&15&14&13&12&11&10&9&8\\\hline
\textbf{2 0}&8&9&10&11&12&13&14&15&0&1&2&3&4&5&6&7\\\hline
\textbf{2 1}&9&8&11&10&13&12&15&14&1&0&3&2&5&4&7&6\\\hline
\textbf{2 2}&10&11&8&9&14&15&12&13&2&3&0&1&6&7&4&5\\\hline
\textbf{2 3}&11&10&9&8&15&14&13&12&3&2&1&0&7&6&5&4\\\hline
\textbf{3 0}&12&13&14&15&8&9&10&11&4&5&6&7&0&1&2&3\\\hline
\textbf{3 1}&13&12&15&14&9&8&11&10&5&4&7&6&1&0&3&2\\\hline
\textbf{3 2}&14&15&12&13&10&11&8&9&6&7&4&5&2&3&0&1\\\hline
\textbf{3 3}&15&14&13&12&11&10&9&8&7&6&5&4&3&2&1&0\\\hline
\end{tabular}
\end{footnotesize}
\label{LS_1_4}
}
\subfigure[Latin Square that removes the singular fade subspace $f_2$]{
\begin{footnotesize}\begin{tabular}{|l|c|c|c|c|c|c|c|c|c|c|c|c|c|c|c|c|}
\hline
&\textbf{0 0}&\textbf{0 1}&\textbf{0 2}&\textbf{0 3}&\textbf{1 0}&\textbf{1 1}&\textbf{1 2}&\textbf{1 3}&\textbf{2 0}&\textbf{2 1}&\textbf{2 2}&\textbf{2 3}&\textbf{3 0}&\textbf{3 1}&\textbf{3 2}&\textbf{3 3}\\\hline
\textbf{0 0}&0&1&2&3&4&5&6&7&8&9&10&11&12&13&15&14\\\hline
\textbf{0 1}&5&4&6&1&0&3&2&10&11&12&14&15&7&8&13&9\\\hline
\textbf{0 2}&9&8&7&10&6&11&14&15&12&13&2&0&1&5&3&4\\\hline
\textbf{0 3}&3&12&11&15&13&14&10&2&4&1&9&7&5&0&6&8\\\hline
\textbf{1 0}&2&0&3&4&5&1&7&6&9&8&11&10&13&12&14&15\\\hline
\textbf{1 1}&6&5&4&7&1&2&0&3&10&14&15&8&9&11&12&13\\\hline
\textbf{1 2}&10&9&8&11&7&12&15&13&14&0&4&1&3&6&2&5\\\hline
\textbf{1 3}&13&14&15&8&9&7&12&11&2&5&6&3&4&10&1&0\\\hline
\textbf{2 0}&1&2&0&5&3&4&8&9&6&7&12&13&14&15&10&11\\\hline
\textbf{2 1}&7&6&5&2&8&9&1&12&13&15&0&14&11&3&4&10\\\hline
\textbf{2 2}&11&10&12&13&14&15&4&0&1&2&3&5&6&9&8&7\\\hline
\textbf{2 3}&14&15&13&9&12&8&11&4&5&3&7&2&10&1&0&6\\\hline
\textbf{3 0}&4&3&1&0&2&6&5&8&7&10&13&9&15&14&11&12\\\hline
\textbf{3 1}&8&7&9&6&10&0&13&14&15&11&1&12&2&4&5&3\\\hline
\textbf{3 2}&12&11&10&14&15&13&3&1&0&4&5&6&8&7&9&2\\\hline
\textbf{3 3}&15&13&14&12&11&10&9&5&3&6&8&4&0&2&7&1\\\hline
\end{tabular}
\end{footnotesize}
\label{LS_2}
}
\subfigure[Latin Square that removes the singular fade subspace $f_3$]{
\begin{footnotesize}\begin{tabular}{|l|c|c|c|c|c|c|c|c|c|c|c|c|c|c|c|c|}
\hline
&\textbf{0 0}&\textbf{0 1}&\textbf{0 2}&\textbf{0 3}&\textbf{1 0}&\textbf{1 1}&\textbf{1 2}&\textbf{1 3}&\textbf{2 0}&\textbf{2 1}&\textbf{2 2}&\textbf{2 3}&\textbf{3 0}&\textbf{3 1}&\textbf{3 2}&\textbf{3 3}\\\hline
\textbf{0 0}&0&4&5&1&6&7&8&9&10&11&12&13&2&14&15&3\\\hline
\textbf{0 1}&3&1&0&2&5&8&4&6&7&9&10&11&12&13&16&15\\\hline
\textbf{0 2}&5&2&1&0&3&9&6&4&8&7&11&10&15&12&14&16\\\hline
\textbf{0 3}&4&8&2&5&0&1&3&10&9&12&13&14&6&16&11&7\\\hline
\textbf{1 0}&2&0&3&4&7&5&9&1&6&13&14&8&16&15&10&11\\\hline
\textbf{1 1}&1&3&7&8&4&6&0&2&11&14&15&16&5&9&12&13\\\hline
\textbf{1 2}&6&5&4&3&1&2&7&0&12&15&16&9&13&10&8&14\\\hline
\textbf{1 3}&7&6&8&9&2&13&15&12&14&16&0&1&3&11&4&10\\\hline
\textbf{2 0}&8&7&6&10&9&14&16&13&15&0&1&2&11&3&5&12\\\hline
\textbf{2 1}&9&10&11&6&8&0&1&15&16&2&3&12&14&4&13&5\\\hline
\textbf{2 2}&10&9&12&11&14&4&5&16&13&6&7&15&0&1&2&8\\\hline
\textbf{2 3}&11&12&9&7&15&16&13&14&1&3&2&0&10&5&6&4\\\hline
\textbf{3 0}&12&11&10&13&16&15&14&3&2&1&4&5&7&8&9&6\\\hline
\textbf{3 1}&13&14&15&16&10&3&12&11&4&5&8&6&1&0&7&9\\\hline
\textbf{3 2}&14&13&16&15&11&12&2&5&0&10&6&4&8&7&3&1\\\hline
\textbf{3 3}&15&16&13&14&12&10&11&8&3&4&5&7&9&6&0&2\\\hline
\end{tabular}
\end{footnotesize}
\label{LS_3}
}

\subfigure[Latin Square that removes the singular fade subspace $f_5$]{
\begin{footnotesize}\begin{tabular}{|l|c|c|c|c|c|c|c|c|c|c|c|c|c|c|c|c|}
\hline
&\textbf{0 0}&\textbf{0 1}&\textbf{0 2}&\textbf{0 3}&\textbf{1 0}&\textbf{1 1}&\textbf{1 2}&\textbf{1 3}&\textbf{2 0}&\textbf{2 1}&\textbf{2 2}&\textbf{2 3}&\textbf{3 0}&\textbf{3 1}&\textbf{3 2}&\textbf{3 3}\\\hline
\textbf{0 0}&0&2&3&1&5&4&8&9&10&11&12&13&6&14&15&7\\\hline
\textbf{0 1}&1&0&2&3&6&7&4&5&8&9&10&11&12&13&14&15\\\hline
\textbf{0 2}&3&1&0&2&7&6&5&4&9&8&11&10&13&15&12&14\\\hline
\textbf{0 3}&4&5&1&0&2&3&6&7&11&10&8&14&15&9&13&12\\\hline
\textbf{1 0}&5&4&6&7&0&1&2&3&12&13&9&15&14&8&10&11\\\hline
\textbf{1 1}&2&3&7&6&4&5&0&1&13&14&15&8&9&12&11&10\\\hline
\textbf{1 2}&6&7&4&5&1&2&3&0&14&15&13&12&10&11&8&9\\\hline
\textbf{1 3}&7&6&5&4&3&8&9&2&15&12&14&0&11&10&1&13\\\hline
\textbf{2 0}&8&9&10&11&12&13&14&15&0&1&2&3&4&5&7&6\\\hline
\textbf{2 1}&9&8&11&10&13&12&15&14&1&0&3&2&5&7&6&4\\\hline
\textbf{2 2}&10&11&12&14&15&0&1&13&2&6&7&5&3&4&9&8\\\hline
\textbf{2 3}&11&10&8&9&14&15&13&12&3&2&0&1&7&6&4&5\\\hline
\textbf{3 0}&12&13&9&15&8&14&10&11&4&3&6&7&0&2&5&1\\\hline
\textbf{3 1}&13&14&15&8&9&11&12&10&7&5&4&6&2&1&0&3\\\hline
\textbf{3 2}&15&12&14&13&11&10&7&6&5&4&1&9&8&3&2&0\\\hline
\textbf{3 3}&14&15&13&12&10&9&11&8&6&7&5&4&1&0&3&2\\\hline
\end{tabular}
\end{footnotesize}
\label{LS_5}
}
\caption{Latin Squares that remove different singular fade subspaces}
\label{LS_set1}
\end{figure*}

\addtocounter{figure}{-1}
\addtocounter{subfigure}{4}
\begin{figure*}
\centering
\subfigure[Latin Square that removes the singular fade subspace $f_6$]{
\begin{footnotesize}\begin{tabular}{|l|c|c|c|c|c|c|c|c|c|c|c|c|c|c|c|c|}
\hline
&\textbf{0 0}&\textbf{0 1}&\textbf{0 2}&\textbf{0 3}&\textbf{1 0}&\textbf{1 1}&\textbf{1 2}&\textbf{1 3}&\textbf{2 0}&\textbf{2 1}&\textbf{2 2}&\textbf{2 3}&\textbf{3 0}&\textbf{3 1}&\textbf{3 2}&\textbf{3 3}\\\hline
\textbf{0 0}&0&1&2&3&8&9&10&11&12&13&14&15&4&5&6&7\\\hline
\textbf{0 1}&8&9&10&11&1&0&3&2&5&4&7&6&12&13&14&15\\\hline
\textbf{0 2}&12&13&14&15&5&4&7&6&2&3&0&1&11&10&9&8\\\hline
\textbf{0 3}&4&5&6&7&12&13&14&15&11&10&9&8&3&2&1&0\\\hline
\textbf{1 0}&1&0&3&2&9&8&11&10&13&12&15&14&5&4&7&6\\\hline
\textbf{1 1}&9&8&11&10&0&1&2&3&4&5&6&7&13&12&15&14\\\hline
\textbf{1 2}&13&12&15&14&4&5&6&7&3&2&1&0&10&11&8&9\\\hline
\textbf{1 3}&5&4&7&6&13&12&15&14&10&11&8&9&2&3&0&1\\\hline
\textbf{2 0}&2&3&0&1&10&11&8&9&14&15&12&13&6&7&4&5\\\hline
\textbf{2 1}&10&11&8&9&3&2&1&0&7&6&5&4&14&15&12&13\\\hline
\textbf{2 2}&14&15&12&13&7&6&5&4&0&1&2&3&9&8&11&10\\\hline
\textbf{2 3}&6&7&4&5&14&15&13&12&9&8&11&10&1&0&3&2\\\hline
\textbf{3 0}&3&2&1&0&11&10&9&8&15&14&13&12&7&6&5&4\\\hline
\textbf{3 1}&11&10&9&8&2&3&0&1&6&7&4&5&15&14&13&12\\\hline
\textbf{3 2}&15&14&13&12&6&7&4&5&1&0&3&2&8&9&10&11\\\hline
\textbf{3 3}&7&6&5&4&15&14&12&13&8&9&10&11&0&1&2&3\\\hline
\end{tabular}
\end{footnotesize}
\label{LS_6}
}
\subfigure[Latin Square that removes the singular fade subspace $f_7$]{
\begin{footnotesize}\begin{tabular}{|l|c|c|c|c|c|c|c|c|c|c|c|c|c|c|c|c|}
\hline
&\textbf{0 0}&\textbf{0 1}&\textbf{0 2}&\textbf{0 3}&\textbf{1 0}&\textbf{1 1}&\textbf{1 2}&\textbf{1 3}&\textbf{2 0}&\textbf{2 1}&\textbf{2 2}&\textbf{2 3}&\textbf{3 0}&\textbf{3 1}&\textbf{3 2}&\textbf{3 3}\\\hline
\textbf{0 0}&0&1&2&3&4&5&6&7&8&9&10&13&12&11&14&15\\\hline
\textbf{0 1}&5&4&6&7&0&8&9&12&13&14&1&2&3&15&10&11\\\hline
\textbf{0 2}&8&9&10&13&14&15&0&1&2&4&12&5&11&6&7&3\\\hline
\textbf{0 3}&12&13&14&9&10&0&15&11&5&1&6&8&7&2&3&4\\\hline
\textbf{1 0}&1&0&3&2&5&4&7&6&9&8&11&12&13&10&15&14\\\hline
\textbf{1 1}&4&5&7&6&1&9&8&13&12&15&0&3&2&14&11&10\\\hline
\textbf{1 2}&9&8&11&12&15&14&1&0&3&5&13&4&10&7&6&2\\\hline
\textbf{1 3}&13&12&15&8&11&1&14&10&4&0&7&9&6&3&2&5\\\hline
\textbf{2 0}&2&3&0&1&6&7&4&5&10&11&8&15&14&9&12&13\\\hline
\textbf{2 1}&7&6&4&5&2&10&11&14&15&12&3&0&1&13&8&9\\\hline
\textbf{2 2}&10&11&8&15&12&13&2&3&0&6&14&7&9&4&5&1\\\hline
\textbf{2 3}&14&15&12&11&8&2&13&9&7&3&4&10&5&1&0&6\\\hline
\textbf{3 0}&3&2&1&0&7&6&5&4&11&10&9&14&15&8&13&12\\\hline
\textbf{3 1}&6&7&5&4&3&11&10&15&14&13&2&1&0&12&9&8\\\hline
\textbf{3 2}&11&10&9&14&13&12&3&2&1&7&15&6&8&5&4&0\\\hline
\textbf{3 3}&15&14&13&10&9&3&12&8&6&2&5&11&4&0&1&7\\\hline
\end{tabular}
\end{footnotesize}
\label{LS_7}
}

\subfigure[Latin Square that removes the singular fade subspace $f_8$]{
\begin{footnotesize}\begin{tabular}{|l|c|c|c|c|c|c|c|c|c|c|c|c|c|c|c|c|}
\hline
&\textbf{0 0}&\textbf{0 1}&\textbf{0 2}&\textbf{0 3}&\textbf{1 0}&\textbf{1 1}&\textbf{1 2}&\textbf{1 3}&\textbf{2 0}&\textbf{2 1}&\textbf{2 2}&\textbf{2 3}&\textbf{3 0}&\textbf{3 1}&\textbf{3 2}&\textbf{3 3}\\\hline
\textbf{0 0}&0&1&2&4&5&9&3&6&7&13&10&14&8&15&11&12\\\hline
\textbf{0 1}&8&12&5&1&0&4&9&13&14&2&3&10&15&11&6&7\\\hline
\textbf{0 2}&5&9&13&14&15&0&4&10&11&8&12&6&1&7&2&3\\\hline
\textbf{0 3}&13&5&9&10&11&14&15&1&2&6&4&0&7&3&12&8\\\hline
\textbf{1 0}&1&0&3&5&4&8&2&7&6&12&11&15&9&14&10&13\\\hline
\textbf{1 1}&9&13&4&0&1&5&8&12&15&3&2&11&14&10&7&6\\\hline
\textbf{1 2}&4&8&12&15&14&1&5&11&10&9&13&7&0&6&3&2\\\hline
\textbf{1 3}&12&4&8&11&10&15&14&0&3&7&5&1&6&2&13&9\\\hline
\textbf{2 0}&2&3&0&6&7&11&1&4&5&15&8&12&10&13&9&14\\\hline
\textbf{2 1}&10&14&7&3&2&6&11&15&12&0&1&8&13&9&4&5\\\hline
\textbf{2 2}&7&11&15&12&13&2&6&8&9&10&14&4&3&5&0&1\\\hline
\textbf{2 3}&15&7&11&8&9&12&13&3&1&4&6&2&5&0&14&10\\\hline
\textbf{3 0}&3&2&1&7&6&10&0&5&4&14&9&13&11&12&8&15\\\hline
\textbf{3 1}&11&15&6&2&3&7&10&14&13&1&0&9&12&8&5&4\\\hline
\textbf{3 2}&6&10&14&13&12&3&7&9&8&11&15&5&2&4&1&0\\\hline
\textbf{3 3}&14&6&10&9&8&13&12&2&0&5&7&3&4&1&15&11\\\hline
\end{tabular}
\end{footnotesize}
\label{LS_8}
}
\subfigure[Latin Square that removes the singular fade subspace $f_9$]{
\begin{footnotesize}\begin{tabular}{|l|c|c|c|c|c|c|c|c|c|c|c|c|c|c|c|c|}
\hline
&\textbf{0 0}&\textbf{0 1}&\textbf{0 2}&\textbf{0 3}&\textbf{1 0}&\textbf{1 1}&\textbf{1 2}&\textbf{1 3}&\textbf{2 0}&\textbf{2 1}&\textbf{2 2}&\textbf{2 3}&\textbf{3 0}&\textbf{3 1}&\textbf{3 2}&\textbf{3 3}\\\hline
\textbf{0 0}&0&4&8&12&1&5&9&11&2&6&10&14&3&7&13&15\\\hline
\textbf{0 1}&5&0&13&3&4&8&14&15&6&9&1&10&7&12&2&11\\\hline
\textbf{0 2}&8&14&2&11&9&15&4&6&10&0&12&7&13&1&5&3\\\hline
\textbf{0 3}&12&10&5&7&13&0&1&2&14&15&6&3&9&11&8&4\\\hline
\textbf{1 0}&1&5&9&13&0&4&8&10&3&7&11&15&2&6&12&14\\\hline
\textbf{1 1}&4&1&12&2&5&9&15&14&7&8&0&11&6&13&3&10\\\hline
\textbf{1 2}&9&15&3&10&8&14&5&7&11&1&13&6&12&0&4&2\\\hline
\textbf{1 3}&13&11&4&6&12&1&0&3&15&14&7&2&8&10&9&5\\\hline
\textbf{2 0}&2&6&10&14&3&7&11&9&0&4&8&12&1&5&15&13\\\hline
\textbf{2 1}&7&2&15&1&6&10&12&13&4&11&3&8&5&14&0&9\\\hline
\textbf{2 2}&10&12&0&9&11&13&6&4&8&2&14&5&15&3&7&1\\\hline
\textbf{2 3}&14&8&7&5&15&2&3&1&12&13&4&0&11&9&10&6\\\hline
\textbf{3 0}&3&7&11&15&2&6&10&8&1&5&9&13&0&4&14&12\\\hline
\textbf{3 1}&6&3&14&0&7&11&13&12&5&10&2&9&4&15&1&8\\\hline
\textbf{3 2}&11&13&1&8&10&12&7&5&9&3&15&4&14&2&6&0\\\hline
\textbf{3 3}&15&9&6&4&14&3&2&0&13&12&5&1&10&8&11&7\\\hline
\end{tabular}
\end{footnotesize}
\label{LS_9}
}
\caption{(Contd.) Latin Squares that remove different singular fade subspaces}
\label{LS_set2}
\end{figure*}
\addtocounter{figure}{-1}
\addtocounter{subfigure}{8}
\begin{figure*}
\centering
\subfigure[Latin Square that removes the singular fade subspace $f_{10}$]{
\begin{footnotesize}\begin{tabular}{|l|c|c|c|c|c|c|c|c|c|c|c|c|c|c|c|c|}
\hline
&\textbf{0 0}&\textbf{0 1}&\textbf{0 2}&\textbf{0 3}&\textbf{1 0}&\textbf{1 1}&\textbf{1 2}&\textbf{1 3}&\textbf{2 0}&\textbf{2 1}&\textbf{2 2}&\textbf{2 3}&\textbf{3 0}&\textbf{3 1}&\textbf{3 2}&\textbf{3 3}\\\hline
\textbf{0 0}&0&3&10&4&8&12&11&14&1&9&13&15&2&5&6&7\\\hline
\textbf{0 1}&9&7&13&15&12&8&14&10&5&0&4&11&6&1&2&3\\\hline
\textbf{0 2}&13&14&3&11&1&2&5&6&9&4&0&7&10&15&8&12\\\hline
\textbf{0 3}&4&9&7&0&5&6&1&2&13&15&11&3&14&10&12&8\\\hline
\textbf{1 0}&1&2&11&5&9&13&10&15&0&8&12&14&3&4&7&6\\\hline
\textbf{1 1}&8&6&12&14&13&9&15&11&4&1&5&10&7&0&3&2\\\hline
\textbf{1 2}&12&15&2&10&0&3&4&7&8&5&1&6&11&14&9&13\\\hline
\textbf{1 3}&5&8&6&1&4&7&0&3&12&14&10&2&15&11&13&9\\\hline
\textbf{2 0}&2&1&8&6&10&14&9&12&3&11&15&13&0&7&4&5\\\hline
\textbf{2 1}&11&5&15&13&14&10&12&8&7&2&6&9&4&3&0&1\\\hline
\textbf{2 2}&15&12&1&9&3&0&7&4&11&6&2&5&8&13&10&14\\\hline
\textbf{2 3}&6&11&5&2&7&4&3&1&15&13&9&0&12&8&14&10\\\hline
\textbf{3 0}&3&0&9&7&11&15&8&13&2&10&14&12&1&6&5&4\\\hline
\textbf{3 1}&10&4&14&12&15&11&13&9&6&3&7&8&5&2&1&0\\\hline
\textbf{3 2}&14&13&0&8&2&1&6&5&10&7&3&4&9&12&11&15\\\hline
\textbf{3 3}&7&10&4&3&6&5&2&0&14&12&8&1&13&9&15&11\\\hline
\end{tabular}
\end{footnotesize}
\label{LS_10}
}

\subfigure[Latin Square that removes the singular fade subspace $f_{12}$]{
\begin{footnotesize}\begin{tabular}{|l|c|c|c|c|c|c|c|c|c|c|c|c|c|c|c|c|}
\hline
&\textbf{0 0}&\textbf{0 1}&\textbf{0 2}&\textbf{0 3}&\textbf{1 0}&\textbf{1 1}&\textbf{1 2}&\textbf{1 3}&\textbf{2 0}&\textbf{2 1}&\textbf{2 2}&\textbf{2 3}&\textbf{3 0}&\textbf{3 1}&\textbf{3 2}&\textbf{3 3}\\\hline
\textbf{0 0}&0&4&1&3&5&9&6&8&10&14&11&13&15&19&16&18\\\hline
\textbf{0 1}&1&2&4&0&6&7&9&5&11&12&14&10&16&17&19&15\\\hline
\textbf{0 2}&2&0&3&4&7&5&8&9&12&10&13&14&17&15&18&19\\\hline
\textbf{0 3}&4&3&2&1&9&8&7&6&14&13&12&11&19&18&17&16\\\hline
\textbf{1 0}&5&9&6&8&0&4&1&3&15&19&16&18&10&14&11&13\\\hline
\textbf{1 1}&6&7&9&5&1&2&4&0&16&17&19&15&11&12&14&10\\\hline
\textbf{1 2}&7&5&8&9&2&0&3&4&17&15&18&19&12&10&13&14\\\hline
\textbf{1 3}&9&8&7&6&4&3&2&1&19&18&17&16&14&13&12&11\\\hline
\textbf{2 0}&10&14&11&13&15&19&16&18&0&4&1&3&5&9&6&8\\\hline
\textbf{2 1}&11&12&14&10&16&17&19&15&1&2&4&0&6&7&9&5\\\hline
\textbf{2 2}&12&10&13&14&17&15&18&19&2&0&3&4&7&5&8&9\\\hline
\textbf{2 3}&14&13&12&11&19&18&17&16&4&3&2&1&9&8&7&6\\\hline
\textbf{3 0}&15&19&16&18&10&14&11&13&5&9&6&8&0&4&1&3\\\hline
\textbf{3 1}&16&17&19&15&11&12&14&10&6&7&9&5&1&2&4&0\\\hline
\textbf{3 2}&17&15&18&19&12&10&13&14&7&5&8&9&2&0&3&4\\\hline
\textbf{3 3}&19&18&17&16&14&13&12&11&9&8&7&6&4&3&2&1\\\hline
\end{tabular}
\end{footnotesize}
\label{LS_12}
}
\caption{(Contd.) Latin Squares that remove different singular fade subspaces}
\label{LS_set3}
\end{figure*}
\end{appendices}
%Consider the case when  The SNR vs BER curve and the SNR vs end-end throughput curves for this case, for a frame length of 256 bits are as shown in Fig. \ref{fig:ber_rician} and Fig. \ref{fig:tput_rician} respectively. From Fig. \ref{fig:ber_rician}. 

%It can be seen from the simulation results presented that when there is a dominant line of sight component, as in the Rician fading scenario, the CNC algorithm performs better than the proposed scheme. The reason for this is as follows: The end to end SNR vs BER as well as the throughput performance depend on the performance during the MA phase as well as the BC phase. When the line of sight component becomes more and more predominant, the performance during the BC phase gets better and better, but the effect of multiple access interference which occurs during the MA phase remains the same. Hence, for the cases when line of sight component is predominant, the performance degradation due to the MA interference predominates over the degradation occurring during the BC phase. The CNC algorithm maximizes the minimum cluster distance and obtains the best distance profile, thereby optimizing the performance during MA phase to the fullest extent possible, at the cost of degraded performance during the BC phase. Hence, the CNC algorithm performs better than the proposed scheme when there is a dominant line of sight component. 

\end{document}